\documentclass[a4paper,11pt]{article}
\usepackage{amsfonts}
\usepackage{amsmath}
\usepackage{amssymb}
\usepackage{colortbl}
\usepackage{hyperref}
\usepackage{cite}
\numberwithin{equation}{section}
\usepackage[left=2cm,right=2cm,top=3.5cm,bottom=3.5cm]{geometry}

\allowdisplaybreaks[1]
\hypersetup{
colorlinks=true,
linkcolor=ceruleanblue,
filecolor=ceruleanblue,      
urlcolor=ceruleanblue,
citecolor=ceruleanblue,
}
\definecolor{ceruleanblue}{rgb}{0.0, 0.2, 0.6}

\date{\today}

\begin{document}

\begin{flushright} {\footnotesize YITP-22-27, IPMU22-0012}  \end{flushright}

\begin{center}
\LARGE{\bf Effective Field Theory of Black Hole Perturbations\\
 with Timelike Scalar Profile: Formulation}
\\[1cm] 

\large{Shinji Mukohyama$^{\,\rm a, \rm b}$ and Vicharit Yingcharoenrat$^{\,\rm b}$}
\\[0.5cm]

\small{
\textit{$^{\rm a}$
Center for Gravitational Physics and Quantum Information, Yukawa Institute for Theoretical Physics, 
\\ Kyoto University, 606-8502, Kyoto, Japan}}
\vspace{.2cm}

\small{
\textit{$^{\rm b}$
Kavli Institute for the Physics and Mathematics of the Universe (WPI), The University of Tokyo Institutes for Advanced Study (UTIAS), The University of Tokyo, Kashiwa, Chiba 277-8583, Japan}}
\vspace{.2cm}
\end{center}

\vspace{0.3cm} 

\begin{abstract}\normalsize
We formulate the Effective Field Theory (EFT) of perturbations within scalar-tensor theories on an inhomogeneous background. The EFT is constructed while keeping a background of a scalar field to be $\textit{timelike}$, which spontaneously breaks the time diffeomorphism. We find a set of consistency relations that are imposed by the invariance of the EFT under the 3d spatial diffeomorphism. This EFT can be generically applied to any inhomogeneous background metric as long as the scalar profile is everywhere timelike. For completeness, we report a dictionary between our EFT parameters to those of Horndeski theories. Finally, we compute background equations for a class of spherically symmetric, static black hole backgrounds, including a stealth Schwarzschild-de Sitter solution. 
\end{abstract}

\vspace{0.3cm} 

\vspace{2cm}

\newpage
{
\hypersetup{linkcolor=black}
\tableofcontents
}

\flushbottom

\vspace{1cm}


\section{Introduction}

Gravitational theories modifying general relativity (GR) have been an attractive and growing topic in recent literature. Such modifications are introduced in order to address mysteries in the universe such as dark energy, dark matter, inflation, big-bang singularity and cosmic magnetic field, which cannot be fully described within GR. Additionally, modifications of GR on short distances may shed light on constructing a consistent theory of quantum gravity. Moreover, modifications of GR can be useful to understand GR itself: deviations from GR predicted by modified gravity theories can be constrained by observations and/or experiments and thus provide an efficient strategy to test the validity of GR in the range of scales relevant to and within the accuracy of the observations/experiments.

Many modified gravity theories have been proposed so far and surely many more will be considered. When one encounters a gravitational theory, there are three check points: ``What are the physical degrees of freedom?'', ``How do they interact?'', and ``What is the regime of validity?'' If two (or more) theories give the same answers to these three questions then they are actually the same even if they look different. This fact suggests the existence of a universal framework. Effective field theories (EFTs) provide such a framework. 

Scalar-tensor theories, described by the spacetime metric and a scalar field, have a relatively long history, initiated by Jordan in 1955~\cite{jordan1955schwerkraft}, Brans-Dicke in 1961~\cite{Brans:1961sx}, Bergmann in 1968~\cite{Bergmann:1968ve}, Wagoner in 1970~\cite{Wagoner:1970vr} and so on. The most general scalar-tensor theory with second-order covariant equations of motion in four-dimensional spacetime was then discovered by Horndeski in 1974~\cite{Horndeski:1974wa}. The so-called degenerate higher-order scalar-tensor (DHOST) theory beyond Horndeski's theory was proposed in 2015~\cite{Langlois:2015cwa} (for a comprehensive review see \cite{Langlois:2018dxi} and references therein). All of them and more should be described universally by an EFT, provided that a symmetry breaking pattern and a background are specified. 

An EFT of scalar-tensor theories in Minkowski and de Sitter backgrounds, called ghost condensation, was developed in \cite{ArkaniHamed:2003uy,Arkani-Hamed:2003juy}, assuming that the time diffeomorphism invariance is spontaneously broken by a timelike derivative of a scalar field and that the spatial diffeomorphism invariance remains unbroken. In particular, the EFT action in the Minkowski background was systematically constructed in the unitary gauge based on the symmetry breaking pattern and then the 4d diffeomorphism invariance was restored by introduction of the Nambu-Goldstone boson. This construction of the EFT action was later extended to a general Friedmann-Lema\^{i}tre-Robertson-Walker (FLRW) background in \cite{Creminelli:2006xe} and served as a precursor of the so called EFT of inflation/dark energy (DE), which we now briefly mention. 

In \cite{Cheung:2007st}, the EFT of inflation was formulated by adopting the slow-roll approximation to rewrite the general EFT action of \cite{Creminelli:2006xe} into a form that is suitable for the study of inflationary cosmological perturbations. 
A further development on the EFT of inflation has led to possibilities of classifying and constraining the higher-order correlation functions~\cite{Creminelli:2005hu,Senatore:2009gt}. Similarly, the EFT of DE \cite{Gubitosi:2012hu} was formulated by rewriting the general EFT action of \cite{Creminelli:2006xe} into a form that is suitable for the study of the structure formation in the presence of a time-evolving scalar field as dark energy. The EFT of DE has been useful to constrain the behavior of scalar-tensor theories in the late-time universe by various cosmological probes and gravitational wave (GW) observations~\footnote{The constraints on dark energy theories after GW170817/GRB170817A can be found in \cite{Creminelli:2017sry,Ezquiaga:2017ekz,Baker:2017hug} and their further constraints from the GW decay into dark energy and from dark energy instabilities induced by GWs were obtained in \cite{Creminelli:2018xsv,Creminelli:2019nok,Creminelli:2019kjy}.}. More recently, the EFT of vector-tensor theories as dark energy was constructed by extending the EFT of gauged ghost condensation \cite{Cheng:2006us} to a FLRW background \cite{Aoki:2021wew}.

In the era of gravitational wave physics/astronomy, one hopes to study scalar-tensor theories as dark energy in the strong gravity regime. For this purpose a black hole immersed in the cosmological background with a time-evolving scalar field would be an ideal setup. It is of course a logical possibility that an EFT describing physics at the cosmological scale breaks down as one approaches the black hole and that physics of the black hole is described by a different EFT. In this case, unfortunately, one cannot hope to acquire any knowledge about dark energy from astrophysical black holes unless one finds a UV completion that can tell connections between the two different EFTs describing different physical regimes. On the other hand, if a single EFT can describe both regimes then one can learn something about dark energy from observations associated with black holes, at least in principle. In the context of scalar-tensor theories with a time-evolving scalar field as dark energy, such an EFT is possible only if the derivative of a scalar field is non-vanishing and timelike all the way from the cosmological asymptotic region to the black hole region. Otherwise, the contours of the scalar field would not be matched between the two spacetime regions smoothly. 

One might wonder if it is possible to find a solution with a regular timelike scalar profile near and on a black hole horizon. This is indeed possible. As an example, let us consider a shift-symmetric k-essence theory $\mathcal{P}(X)$ with $X=\partial^{\mu}\Phi\partial_{\mu}\Phi$ and assume that $\mathcal{P}'(X_0)=0$ for ${}^{\exists} X_0 = const. < 0$. Consider any black hole solution $g_{\mu\nu}$ in GR and choose an arbitrary spacelike hypersurface $\Sigma$ that penetrates the black hole horizon. In a neighborhood of $\Sigma$, one can find the Gaussian normal coordinate system by considering a congruence of timelike geodesics orthogonal to $\Sigma$. By setting $\Phi=\sqrt{-X_0}\tau$, where $\tau$ is the time in the Gaussian normal coordinate system, and shifting the bare cosmological constant by the amount $\mathcal{P}(X_0)$, the GR solution survives as a solution in the k-essence theory minimally coupled to GR. In order to see that this is indeed the case, one only needs to note that the stress-energy tensor of the k-esssence field is $T^{\Phi}_{\mu\nu} = -2\mathcal{P}'(X) \partial_{\mu}\Phi \partial_{\nu}\Phi + \mathcal{P}(X) g_{\mu\nu} = \mathcal{P}(X_0)g_{\mu\nu}$ for $X=X_0$. This kind of solutions, where the spacetime metric is a solution in GR and where the scalar field profile is non-trivial, is often called stealth solutions. This construction of stealth solutions locally works without assuming any symmetries, but breaks down when the congruence of geodesics forms caustics. The first stealth black hole solution was found in \cite{Mukohyama:2005rw} by adopting a globally well-behaved Gaussian normal coordinate system called the Lema\^{i}tre reference frame in the Schwarzschild spacetime. The stealth Schwarzschild-(A)dS solution was then found in Horndeski theory \cite{Babichev:2013cya,Kobayashi:2014eva} and in DHOST theory \cite{BenAchour:2018dap,Motohashi:2019sen}. The stealth Kerr-(A)dS solution was then found in DHOST theory \cite{Charmousis:2019vnf}. One can also find non-stealth solutions by using disformal transformations as a solution-generating method~\cite{BenAchour:2020wiw,BenAchour:2020fgy,Anson:2020trg}. 

While perturbations around most of those stealth solutions in DHOST theories are infinitely strongly coupled \cite{deRham:2019gha}, one can introduce higher derivative terms to render them weakly coupled in a theoretically controlled manner without significantly changing properties of the metric at astrophysical scales. This mechanism, dubbed scordatura, was implemented already in the EFT of ghost condensation \cite{ArkaniHamed:2003uy,Mukohyama:2005rw} and recently revisited in a broader context \cite{Motohashi:2019ymr,Gorji:2020bfl,Gorji:2021isn}. The scordatura mechanism works as far as the derivative of the scalar field is non-vanishing and timelike but does not work for a spacelike case, and this fact provides yet another motivation for the timelike scalar profile. 

The purpose of the present paper is to construct the EFT of black hole perturbations in the context of scalar-tensor theories with a \textit{timelike} scalar profile~\footnote{The no-hair theorems for scalar-tensor theories \cite{Hui:2012qt,Saravani:2019xwx,Babichev:2016rlq,Antoniou:2017acq,Babichev:2013cya} do not apply to our case due to a non-static scalar configuration around stationary black hole spacetimes. For a similar EFT with a static, spacelike scalar profile see \cite{Franciolini:2018uyq}.}. This EFT reduces to the EFT of DE in the asymptotic region, can be applied to both stealth and non-stealth solutions, and naturally accommodates the scordatura mechanism for the former. Furthermore, the methodology developed in this paper does not assume any symmetries and can be applied to both static/stationary and dynamical backgrounds while the concrete computations are performed for a class of static, spherically symmetric black holes. 

The rest of the present paper is organized as follows. In Section~\ref{sec2:EFT_inhomo} we develop the general methodology to construct an EFT of scalar-tensor theories around an inhomogeneous background metric with a timelike scalar profile. In Section~\ref{sec:Dictionary} we then give a dictionary between the EFT and a concrete theory, focusing for simplicity on the Horndeski theory. In Section~\ref{sec:bg_tadpole}, for simplicity we restrict our considerations to a class of static, spherically symmetric backgrounds and obtain the background equations of motion from the EFT action. Section~\ref{sec:conclusions} is devoted to a summary of the paper and future directions.

\section{EFT on inhomogeneous background}\label{sec2:EFT_inhomo}
\subsection{Action in unitary gauge}
In this Section we will explain how to construct an EFT of perturbations in the unitary gauge for the scalar-tensor theories on a background metric that is not necessarily spatially homogeneous. The key point here is that we want to write down the EFT in the presence of a time-dependent background of a scalar field, $\bar{\Phi}(\tau)$, where $\tau$ defines a time direction. This is indeed the main difference between our EFT and the one constructed in \cite{Franciolini:2018uyq} where they assumed the background scalar field to be spacelike. Note that in the language of field theory the background $\bar{\Phi}(\tau)$ spontaneously breaks $\tau$-diffeomorphism. 

Let us now introduce the unitary gauge formalism. In this gauge one simply sets a perturbation of $\Phi$ to zero, $\delta \Phi = 0$; all perturbations are inside the metric. In doing so, the scalar field defines a preferred ($\Phi = const.$) time-slicing whose unit normal vector can be defined by
\begin{align}\label{eq:normal_vec}
n_\mu = - \frac{\partial_\mu \Phi}{\sqrt{-(\partial \Phi)^2}}  \rightarrow - \frac{\dot{\bar{\Phi}}\delta_\mu^\tau}{\sqrt{-g^{\tau\tau}\dot{\bar{\Phi}}^2}}  \;, 
\end{align}
with $n_\mu n^\mu = -1$. The right hand side of the arrow represents the expression in the unitary gauge and a dot denotes a derivative with respect to $\tau$. One can further choose $\bar{\Phi} \propto \tau$, but here we would like to keep $\bar{\Phi}$ as an arbitrary function of $\tau$.

It is then convenient to introduce the ADM decomposition where the metric can be put in the following form:
\begin{align}\label{eq:ADM_metric}
ds^2 = -N^2 d\tau^2 + h_{ij}(dx^i + N^i d\tau)(dx^j + N^j d\tau) \;,
\end{align}
where $h_{\mu\nu} \equiv g_{\mu\nu} + n_\mu n_\nu$ is an induced metric on the constant-$\tau$ hypersurface, $N$ is a lapse function and $N^i$ is a shift vector.  For later use, written in terms of the ADM variables the components of the metric and its inverse are explicitly given by
\begin{align}\label{eq:metric_ADM}
g_{\tau\tau} = -N^2 + h_{ij}N^i N^j \;, \quad g_{i\tau} = h_{ij}N^j \equiv N_i \;, \quad g_{ij} = h_{ij} \;, 
\end{align}
and 
\begin{align}\label{eq:inverse_metric_ADM}
g^{\tau\tau} = -\frac{1}{N^2} \;, \quad g^{i\tau} = \frac{N^i}{N^2} \;, \quad g^{ij} = h^{ij} - \frac{N^iN^j}{N^2} \;.
\end{align}
Note that the spatial index $i$ is raised or lowered by the induced metric $h_{ij}$. Moreover, using the projection along and orthogonal to the hypersurface, we define the extrinsic curvature,
\begin{align}
K_{\mu\nu} \equiv h_\mu^\sigma \nabla_\sigma n_\nu \;,
\end{align}
with $\nabla_\mu$ being the 4d covariant derivative. Specifically, the spatial components of the extrinsic curvature and its trace can be explicitly written as
\begin{align}
K_{ij} = \frac{1}{2N}(\dot{h}_{ij} - D_iN_j - D_jN_i) \;, \quad K = h^{ij}K_{ij} \;,
\end{align} 
where $D_i$ refers to the covariant spatial derivative constructed from the induced metric. Furthermore, the 3d curvature ${}^{(3)}\!R$ can be computed using the induced metric as in the usual way. 
 
We see that on top of the 4d covariant terms such as the Ricci scalar, $\tilde{R}$, the action in the unitary gauge can also depend on all the geometrical quantities that are invariant under 3d diffeomorphism, for example, $N$, $K_{\mu\nu}$ and ${}^{(3)}\!R_{\mu\nu}$. Besides, due to the spontaneous symmetry breaking of $\tau$-diffeomorphism the explicit $\tau$-dependent function is allowed in the Lagrangian.  Given all the possibilities the most generic action in the unitary gauge is thus given by (see Appendix A of \cite{Cheung:2007st})
\begin{align}\label{eq:EFT_generic}
S = \int d^4x \sqrt{-g}~F(\tilde{R}_{\mu\nu\alpha\beta},g^{\tau\tau},K_{\mu\nu},\nabla_\nu,\tau) \;,
\end{align}
where $\tilde{R}_{\mu\nu\alpha\beta}$ is the 4d Riemann tensor of the metric $g_{\mu\nu}$. The action above can be applied to generic background geometries without any symmetries. However, this is not yet the EFT action that we would like to achieve; it is just an arbitrary function of all possible quantities one can write down in the unitary gauge. 

The story so far follows what has been done in the case of the EFT of inflation/DE \cite{Cheung:2007st,Gubitosi:2012hu}. However, if one defines the building blocks of the EFT to be 
\begin{align}\label{eq:pert}
\delta g^{\tau\tau} \equiv g^{\tau\tau} - \bar{g}^{\tau\tau}(\tau,\vec{x})\;, \quad \delta K^\mu_\nu = K^\mu_\nu - \bar{K}^\mu_\nu(\tau,\vec{x})\;, \quad \delta {}^{(3)}\!R^\mu_\nu \equiv {}^{(3)}\!R^\mu_\nu - {}^{(3)}\!\bar{R}^\mu_\nu(\tau,\vec{x}) \;,
\end{align}
where the background quantities are denoted with the bar and $\vec{x}$ denotes spatial coordinates, and if one naively writes down the EFT action as, for instance,
\begin{align}\label{eq:action_break_3d}
S = \int d^4 x \sqrt{-g}\bigg[\frac{M_\star^2}{2}f\, R + \Lambda - c\, g^{\tau\tau} + \frac{1}{2}m_2^4(\delta g^{\tau\tau})^2  + \frac{1}{2}M_2^2 \delta K^2 + \frac{1}{2}\mu_1^2\delta g^{\tau\tau} \delta {}^{(3)}\!R + \ldots \bigg] \;,
\end{align}
(see (\ref{eqn:def-R}) for the definition of $R$) then this naive EFT action is not manifestly invariant under the 3d diffeomorphism. The reason is that in general the background part of each quantity in (\ref{eq:pert}) depends on the spatial coordinates and thus the perturbative quantities defined in (\ref{eq:pert}) do not transform as naively expected under the spatial diffeomorphism. In particular, a naive ``scalar'' made of them by contracting spacetime indices does not actually transform as a scalar under the spatial diffeomorphism. Therefore, the action (\ref{eq:action_break_3d}) breaks the 3d diffeomorphism explicitly unless additional conditions are imposed.

We will see below that the only way to write down the EFT action without explicitly breaking the spatial diffeomorphism is to impose a certain consistency condition among the EFT parameters. In the next Section we will then start with the unitary gauge action (\ref{eq:EFT_generic}) and explicitly write down a condition under which the EFT action is invariant under the 3d diffeomorphism.

\subsection{Expansion around an inhomogeneous background}

As we pointed out in the previous Section, the naive EFT expansion (\ref{eq:action_break_3d}) does not directly give rise to a consistent EFT in which the 3d diffeomorphism is preserved. In this Section we start with the most generic action in the unitary gauge (\ref{eq:EFT_generic}) and expand it around the non-trivial $\vec{x}$-dependent background quantities. Finally, we will see that a set of consistency relations among the coefficients in (\ref{eq:action_break_3d}) must be satisfied for the expanded action (\ref{eq:action_break_3d}) to be descendant from the general action (\ref{eq:EFT_generic}) that is manifestly invariant under the 3d diffeomorphism. 

Recall the most generic action in the unitary gauge (\ref{eq:EFT_generic}),
\begin{align}
S = \int d^4x \sqrt{-g}~F(R_{\mu\nu\alpha\beta},g^{\tau\tau},K_{\mu\nu},\nabla_\nu,\tau) \;.
\end{align}
By construction, this action is manifestly invariant under the 3d diffeomorphism and is applicable for any generic background metric. Given the perturbations and the non-trivial $\vec{x}$-dependent background values, defined in (\ref{eq:pert}), we then Taylor expand the action above up to second order as
\begin{align}\label{eq:EFT_taylor}
S = \int d^4x \sqrt{-g}~\bigg[&\bar{F} + \bar{F}_{g^{\tau\tau}} \delta g^{\tau\tau} + \bar{F}_{K}\delta K + \bar{F}_{\sigma^\mu_\nu}\delta \sigma^\mu_\nu + \bar{F}_{{}^{(3)}\!R}\delta {}^{(3)}\!R +  \bar{F}_{r^\mu_\nu}\delta r^\mu_\nu + \frac{1}{2}\bar{F}_{g^{\tau\tau}g^{\tau\tau}}(\delta g^{\tau\tau})^2 \nonumber \\  & + \bar{F}_{g^{\tau\tau}K} \delta g^{\tau\tau}\delta K + \bar{F}_{g^{\tau\tau} \sigma^\mu_\nu}\delta \sigma^\mu_\nu \delta g^{\tau\tau}  + \bar{F}_{g^{\tau\tau} r^\mu_\nu}\delta r^\mu_\nu \delta g^{\tau\tau} + \bar{F}_{g^{\tau\tau}{}^{(3)}\!R}\delta g^{\tau\tau} \delta {}^{(3)}\!R  \nonumber \\
&   + \frac{1}{2}\bar{F}_{KK} \delta K^2 + \bar{F}_{K \sigma^\mu_\nu}\delta K \delta \sigma^\mu_\nu + \bar{F}_{K{}^{(3)}\!R}\delta K \delta {}^{(3)}\!R + \bar{F}_{K r^\mu_\nu}\delta K \delta r^\mu_\nu    \nonumber \\ 
&  + \frac{1}{2} \bar{F}_{{}^{(3)}\!R{}^{(3)}\!R}\delta {}^{(3)}\!R^2   + \bar{F}_{{}^{(3)}\!R \sigma^\mu_\nu} \delta {}^{(3)}\!R \delta \sigma^\mu_\nu  +   \bar{F}_{{}^{(3)}\!R r^\mu_\nu}\delta {}^{(3)}\!R \delta r^\mu_\nu+\frac{1}{2}\bar{F}_{r^2 }\delta r^\mu_\nu \delta r^\nu_\mu   \nonumber \\
& + \frac{1}{2}\bar{F}_{\sigma^2} \delta \sigma^\mu_\nu \delta \sigma^\nu_\mu + \bar{F}_{\sigma r }\delta \sigma^\mu_\nu \delta r^\nu_\mu + \ldots  \bigg] \;,
\end{align}
where $\sigma_{\mu\nu} \equiv K_{\mu\nu} - K h_{\mu\nu}/3$ and $r_{\mu\nu} \equiv {}^{(3)}\!R_{\mu\nu} - {}^{(3)}\!R h_{\mu\nu}/3$. Notice that the barred quantities denote the Taylor coefficients evaluated on the background: $\bar{F}_X \equiv (\partial F/\partial X )_{\rm BG}$. For completeness, we also included the terms linear in $\delta \sigma_{\mu\nu}$ and $\delta r_{\mu\nu}$ since generally those terms do not vanish~\footnote{For GR with a minimally coupled scalar field on a flat FLRW metric the only non-vanishing linear term is the $\delta K$ term.}. Although it is true that the term linear in $\delta K$ can be absorbed into the other terms, e.g.~$\bar{F}$, by performing an integration by parts to $K = \nabla_\mu n^\nu$, here we would like to keep this term explicitly since the matching with the EFT action we are going to write down will become more transparent. 

As one would expect, the translational symmetries in $\vec{x}$-directions of each term in (\ref{eq:EFT_taylor}) are broken since their Taylor coefficients evaluated on the background in general depend on the spatial coordinates $\vec{x} = \{x^i\}$. However, since the whole Taylor series in (\ref{eq:EFT_taylor}) are descendant from the general action (\ref{eq:EFT_generic}) that is invariant under the spatial diffeomorphism, there must be a set of relations among the coefficients which eventually ensures that the action (\ref{eq:EFT_taylor}) is also invariant under the 3d diffeomorphism.

The consistency relations can be obtained by using the fact that those Taylor coefficients in (\ref{eq:EFT_taylor}) may depend implicitly on the spatial coordinates $\vec{x}$ through the background quantities. On the other hand, they may depend on $\tau$ not only implicitly but also explicitly. (The general action (\ref{eq:EFT_generic}) in the unitary gauge depends on $\tau$ explicitly but does not depend on $\vec{x}$ explicitly.) We then apply the chain rule to $\bar{F}(\tau,\vec{x})$,
\begin{align}
\frac{\partial}{\partial\tau}\bar{F}(\tau,\vec{x}) &= \frac{d}{d\tau} F(g^{\tau\tau},K,{}^{(3)}\!R,\tau) \bigg|_{BG} \nonumber \\
&= \bar{F}_{\tau} + \bar{F}_{g^{\tau\tau}}\frac{\partial \bar{g}^{\tau\tau}}{\partial\tau} + \bar{F}_{K}\frac{\partial \bar{K}}{\partial\tau} + \bar{F}_{{}^{(3)}\!R} \frac{\partial {}^{(3)}\!\bar{R}}{\partial\tau}+ \bar{F}_{\sigma^\mu_\nu} \frac{\partial \bar{\sigma}^\mu_\nu}{\partial\tau} + \bar{F}_{r^\mu_\nu} \frac{\partial \bar{r}^\mu_\nu}{\partial\tau}  + \ldots \label{eq:con1}\;, \\
\frac{\partial}{\partial x^i}\bar{F}(\tau,\vec{x}) &= \frac{d}{dx^i} F(g^{\tau\tau},K,{}^{(3)}\!R,\tau) \bigg|_{BG} \nonumber \\
&= \bar{F}_{g^{\tau\tau}}\frac{\partial \bar{g}^{\tau\tau}}{\partial x^i} + \bar{F}_{K}\frac{\partial \bar{K}}{\partial x^i} + \bar{F}_{\sigma^\mu_\nu} \frac{\partial \bar{\sigma}^\mu_\nu}{\partial x^i} + \bar{F}_{{}^{(3)}\!R} \frac{\partial {}^{(3)}\!\bar{R}}{\partial x^i} + \bar{F}_{r^\mu_\nu} \frac{\partial \bar{r}^\mu_\nu}{\partial x^i} + \ldots \label{eq:con2}\;,
\end{align}
where $\ldots$ refers to contributions of the terms higher-order in derivatives. The appearance of $\bar{F}_{\tau}$ on the right hand side of (\ref{eq:con1}) is due to the fact that there is an explicit $\tau$-dependence in $\bar{F}$. Since $\bar{F}_{\tau}$ does not show up in the EFT action in the unitary gauge, one can always find $\bar{F}_{\tau}$ that satisfies eq.(\ref{eq:con1}) for given values of the EFT parameters. Therefore, the relations involving the $\tau$-derivatives such as eq.(\ref{eq:con1}) do not give rise to a non-trivial relation among the EFT parameters. On the other hand, eq.(\ref{eq:con2}) does not involve $\bar{F}_{x^i}$ ($\bar{F}$ does not explicitly depend on $x^i$), and thus it gives us a non-trivial relation among the Taylor coefficients. Therefore, in what follows we can safely restrict our considerations to the relations involving the $x^i$-derivatives. In particular, the chain rule applied to $\bar{F}_{g^{\tau\tau}}$, $\bar{F}_{K}$, and $\bar{F}_{{}^{(3)}\!R}$ gives
\begin{align}
 \frac{\partial}{\partial x^i} \bar{F}_{g^{\tau\tau}} &= \bar{F}_{g^{\tau\tau}g^{\tau\tau}}\frac{\partial \bar{g}^{\tau\tau}}{\partial x^i} + \bar{F}_{g^{\tau\tau}K}\frac{\partial \bar{K}}{\partial x^i} + \bar{F}_{g^{\tau\tau}\sigma^\mu_\nu}\frac{\partial \bar{\sigma}^\mu_\nu}{\partial x^i} + \bar{F}_{g^{\tau\tau}{}^{(3)}\!R} \frac{\partial {}^{(3)}\!\bar{R} }{\partial x^i}+ \bar{F}_{g^{\tau\tau}r^\mu_\nu}\frac{\partial \bar{r}^\mu_\nu}{\partial x^i} + \ldots \;, \label{eq:con4}\\
\frac{\partial}{\partial x^i}\bar{F}_{K} &= \bar{F}_{g^{\tau\tau}K}\frac{\partial \bar{g}^{\tau\tau}}{\partial x^i} + \bar{F}_{KK}\frac{\partial \bar{K}}{\partial x^i} + \bar{F}_{K\sigma^\mu_\nu}\frac{\partial \bar{\sigma}^\mu_\nu}{\partial x^i}  + \bar{F}_{K{}^{(3)}\!R}\frac{\partial {}^{(3)}\!\bar{R}}{\partial x^i} + \bar{F}_{K r^\mu_\nu} \frac{\partial \bar{r}^\mu_\nu}{\partial x^i} +  \ldots \;, \label{eq:con6} \\
\frac{\partial}{\partial x^i}\bar{F}_{\sigma^\mu_\nu} &= \bar{F}_{g^{\tau\tau} \sigma^\mu_\nu} \frac{\partial \bar{g}^{\tau\tau}}{\partial x^i} + \bar{F}_{K \sigma^\mu_\nu} \frac{\partial \bar{K} }{\partial x^i} + \bar{F}_{\sigma^2} \frac{\partial  \bar{\sigma}_\mu^\nu}{\partial x^i} + \bar{F}_{{}^{(3)}\!R \sigma^\mu_\nu} \frac{\partial {}^{(3)}\!\bar{R}}{\partial x^i} + \bar{F}_{\sigma r} \frac{\partial \bar{r}^\nu_\mu}{\partial x^i}  + \ldots \;,  \label{eq:con9}  
\end{align}
where again $\ldots$ refers to contributions of higher derivative terms. Moreover, one can easily apply the chain rule to other terms e.g.~$\bar{F}_{\sigma}$, $\bar{F}_{r}$ and $\bar{F}_{g^{\tau\tau}g^{\tau\tau}}$. We now impose the consistency relations (\ref{eq:con2})--(\ref{eq:con9}) (and similar ones for other terms) on the action (\ref{eq:EFT_taylor}) to ensure that the unbroken spatial diffeomorphism is preserved up to the order of interest (e.g. the leading order). Note that the same technique was extremely useful to construct the EFT of vector-tensor theories \cite{Aoki:2021wew} in a way that one keeps the action invariant under 3d diffeomorphism.  We will use these relations in the next Subsection.

Before closing this Subsection let us now discuss a typical scale of each term in (\ref{eq:EFT_taylor}). To do so, we define an energy scale $E$ in a region of interest specified by coordinates $\tau$ and $\vec{x}$, as 
\begin{align}\label{eq:energy_E}
E \equiv {\rm max}\{|{}^{(3)}\!\bar{R}|^{1/2} , |{}^{(3)}\!\bar{R}^{\mu}_{\nu}{}^{(3)}\!\bar{R}^{\nu}_{\mu}|^{1/4} , |\bar{K}| , |\bar{K}^{\mu}_{\nu}\bar{K}^{\nu}_{\mu}|^{1/2} , |{}^{(3)}\!\bar{R}^{\mu}_{\nu}\bar{K}^{\nu}_{\mu}|^{1/3}\} \;.
 \end{align}
This energy scale captures the relevant scales of the background metric and has nothing to do with the theories we are interested in. Note that in the case of a flat FLRW metric with a reasonable equation of state for matter this energy scale essentially coincides with the Hubble expansion rate. We also introduce the bare Planck mass $M_\star$  so that the coefficient of ${}^{(3)}\! R$ in $F$ is of $\mathcal{O}(M_\star^2)$. Then, the typical orders of the Taylor coefficients read, for instance, as
 \begin{align}
 \bar{F} &\sim \mathcal{O}(M_\star^2 E^2) \;, \quad  \bar{F}_{K} \sim \bar{F}_{\sigma}  \sim \mathcal{O}(M_\star^2 E) \;, \quad \bar{F}_{{}^{(3)}\!R} \sim \bar{F}_{r}  \sim \mathcal{O}(M_\star^2) \;, \label{eq:scaling1} \\ 
 \bar{F}_{g^{\tau\tau}g^{\tau\tau}} &\sim \mathcal{O}(M_\star^2 E^2) \;, \quad \bar{F}_{g^{\tau\tau}K} \sim \mathcal{O}(M_\star^2 E) \;, \quad \bar{F}_{KK} \sim \bar{F}_{\sigma^2} \sim  \bar{F}_{g^{\tau\tau}{}^{(3)}\!R} \sim \mathcal{O}(M_\star^2) \;. \label{eq:scaling2}
  \end{align}
We then assume that the typical scale for the $\tau$-dependent background of $\Phi$ is $\mu$, which is the scale at which Lorentz invariance is spontaneously broken~\footnote{In the case of EFT of DE this scale $\mu$ is nothing but $\Lambda_2$, that is of the order of $10^{-3}~{\rm eV}$ at the present time.}. In addition, the higher derivative terms will be further suppressed by the cut-off scale $\Lambda_\star$, for example,
 \begin{align}
 \bar{F}_{K{}^{(3)}\!R} \sim \bar{F}_{r{}^{(3)}\!R} \sim \mathcal{O}(M_\star^2 E \Lambda_\star^{-2}) \;, \quad  \bar{F}_{{}^{(3)}\!R{}^{(3)}\!R} \sim \bar{F}_{r^2} \sim \mathcal{O}(M_\star^2 \Lambda_\star^{-2}) \;. \label{eq:scaling3}
 \end{align} 
It is reasonable to further assume that $\mu \gg \Lambda_\star$ and $E < \Lambda_\star$.  Note that written in terms of $M_\star$ and $E$ the scales $\mu$ and $\Lambda_\star$ are of the order of $(M_\star E)^{1/2}$ and $(M_\star E^2)^{1/3}$ respectively.

 \subsection{EFT action}
 
In the last Subsection we have obtained the unitary gauge action expanded around an inhomogeneous background geometry, eq.(\ref{eq:EFT_taylor}). In addition, we found the consistency relations among the Taylor coefficients to make sure that the 3d diffeomorphism is unbroken. 

In this Subsection we write down the most general EFT action up to second order in perturbations. At first glance this EFT action does not appear to be invariant under the 3d diffeomorphism due to the presence of the spatially-dependent functions. This is not surprising since the perturbations such as $\delta K$ and $\delta {}^{(3)}\!R$ are not building blocks of the EFT; the background part of each of them depends on the spatial coordinates $\vec{x}$.  Note that this also happened to the action (\ref{eq:EFT_taylor}) where each term of the Taylor expansion is not invariant under the 3d diffeomorphism. However, as we have pointed out in the previous Subsection the residual gauge symmetries can be reassured if and only if the consistency relations (\ref{eq:con2})--(\ref{eq:con9}) (and similar relations for other terms) hold.  Here, after we write down the EFT action we will then impose the consistency relations among the EFT parameters.
We write the most general EFT action up to second order in perturbations: 
\begin{align}\label{eq:EFT}
S = \int d^4x \sqrt{-g} \bigg[&\frac{M_\star^2}{2}f(y)R - \Lambda(y) - c(y)g^{\tau\tau} - \beta(y) K  - \alpha^\mu_{\nu}(y)\sigma^{\nu}_\mu - \gamma^\mu_{\nu}(y)r^{\nu}_\mu + \frac{1}{2} m_2^4(y) (\delta g^{\tau\tau})^2 \nonumber \\ 
&  + \frac{1}{2} M_1^3(y) \delta g^{\tau\tau} \delta K + \frac{1}{2} M_2^2(y) \delta K^2 + \frac{1}{2} M_3^2(y) \delta K^\mu_\nu \delta K^\nu_\mu + \frac{1}{2}M_4(y) \delta K \delta {}^{(3)}\!R \nonumber \\
& + \frac{1}{2}M_5(y) \delta K^\mu_\nu \delta {}^{(3)}\!R^\nu_\mu + \frac{1}{2}\mu_1^2(y) \delta g^{\tau\tau} \delta {}^{(3)}\!R + \frac{1}{2}\mu_2(y)\delta {}^{(3)}\!R^2 + \frac{1}{2} \mu_3(y) \delta {}^{(3)}\!R^\mu_\nu \delta {}^{(3)}\!R^\nu_\mu \nonumber \\
&+ \frac{1}{2} \lambda_1(y)^\nu_\mu \delta g^{\tau\tau} \delta K^\mu_\nu + \frac{1}{2} \lambda_2(y)^\nu_\mu \delta g^{\tau\tau} \delta {}^{(3)}\!R^\mu_\nu +  \frac{1}{2} \lambda_3(y)^\nu_\mu \delta K \delta K^\mu_\nu +  \frac{1}{2} \lambda_4(y)^\nu_\mu \delta K \delta {}^{(3)}\!R^\mu_\nu \nonumber \\ 
& + \frac{1}{2} \lambda_5(y)^\nu_\mu \delta {}^{(3)}\!R \delta K^\mu_\nu + \frac{1}{2} \lambda_6(y)^\nu_\mu \delta {}^{(3)}\!R \delta {}^{(3)}\!R^\mu_\nu + \ldots \bigg] \;,
\end{align}
where the functions $f(y)$, $\Lambda(y)$, $c(y)$, $\beta(y)$, $\alpha^\mu_\nu(y)$ and $\gamma^\mu_\nu(y)$ will be fixed by the background equations of motion (see Section~\ref{sec:bg_tadpole}), and $y = \{\tau, \vec{x}\}$~\footnote{One should keep in mind that all the spatially-dependent terms are coming from the background quantities such as $\bar{K}$ and ${}^{(3)}\!\bar{R}$.}. For simplicity, we work with the quantity $R$ that corresponds to the 4d Ricci scalar $\tilde{R} \equiv g^{\mu\nu}\tilde{R}_{\mu\nu}$ (with $\tilde{R}_{\mu\nu}\equiv g^{\rho\sigma}\tilde{R}_{\mu\rho\nu\sigma}$) after subtracting the divergence term,
\begin{align} \label{eqn:def-R}
R &\equiv {}^{(3)}\!R + K_{\mu\nu}K^{\mu\nu} - K^2 = \tilde{R} + 2\nabla_\mu(n^\nu\nabla_\nu n^\mu - K n^\mu)\;.
\end{align}
Of course, this boundary term will not affect the equations of motion, but it in fact gives a contribution regarding the dictionary between the EFT action and the covariant theories. Besides, if one wants to compare our EFT action (\ref{eq:EFT}) to the one of inflation/DE in which the original 4d Ricci scalar $\tilde{R}$ was used, one needs to take into account this boundary term. Throughout this paper we use the quantity $R$ defined above. Notice also that in (\ref{eq:EFT}) we kept the term linear in $K$, unlike the EFT of inflation/DE where one performs an integration by parts to $K = \nabla_\mu n^\mu$ and absorbs the $\beta$-term into the other terms in the EFT action. It is important to point out that the $\lambda$'s terms in (\ref{eq:EFT}) were not present in the EFT of inflation/DE since they are redundant operators~\footnote{This is due to the fact that both $K_{\mu\nu}$ and ${}^{(3)}\!R_{\mu\nu}$ on the FLRW background are proportional to the induced metric $h_{\mu\nu}$.}. 

The EFT parameters in (\ref{eq:EFT}) and the Taylor coefficients in (\ref{eq:EFT_taylor}) are related to each other via
\begin{align}
M_\star^2 f(y) &= 2 \bar{F}_{{}^{(3)}\!R} \;, \quad \Lambda(y) = - \bar{F} + \bar{F}_{K} \bar{K}  + \bar{F}_{{}^{(3)}\!R} \bigg({}^{(3)}\!\bar{R} + \frac{2}{3} \bar{K}^2 - \bar{\sigma}^\mu_\nu \bar{\sigma}^\nu_\mu \bigg) + \bar{F}_{g^{\tau\tau}} \bar{g}^{\tau\tau} + \bar{F}_{\sigma^\mu_\nu}\bar{\sigma}^\mu_\nu + \bar{F}_{r^\mu_\nu}\bar{r}^\mu_\nu\;, \nonumber \\
c(y) &= - \bar{F}_{g^{\tau\tau}} \;, \quad \beta(y) = - \bar{F}_K - \frac{4}{3} \bar{F}_{{}^{(3)}\!R} \bar{K} \;, \quad \alpha^\mu_\nu(y) = -\bar{F}_{\sigma^\nu_\mu} + 2 \bar{F}_{{}^{(3)}\!R}\bar{\sigma}^\mu_\nu \;, \quad \gamma^\mu_\nu(y) = -\bar{F}_{r^\nu_\mu} \;, \nonumber \\
m_2^4(y) &= \bar{F}_{g^{\tau\tau}g^{\tau\tau}} \;, \quad M_1^3(y) = 2\bar{F}_{g^{\tau\tau}K} - \frac{2}{3}h^\mu_\nu \bar{F}_{g^{\tau\tau} \sigma^\mu_\nu}  \;, \quad M_2^2 = \bar{F}_{KK} - \frac{1}{3}\bar{F}_{\sigma^2} + 2\bar{F}_{{}^{(3)}\!R} - \frac{2}{3} h^\mu_\nu \bar{F}_{K \sigma^\mu_\nu} \nonumber  \\
M_3^2(y) &= \bar{F}_{\sigma^2} - 2\bar{F}_{{}^{(3)}\!R} \;, \quad  M_4(y) = 2\bar{F}_{K{}^{(3)}\!R} - \frac{2}{3}\bar{F}_{\sigma r} - \frac{2}{3} h^\mu_\nu \bar{F}_{K r^\mu_\nu} - \frac{2}{3} h^\mu_\nu \bar{F}_{{}^{(3)}\!R \sigma^\mu_\nu} \;, \label{eq:EFT_parameters} \\
M_5(y) &= 2 \bar{F}_{\sigma r} \;, \quad \mu_1^2(y) = 2\bar{F}_{g^{\tau\tau}{}^{(3)}\!R} - \frac{2}{3} h^\mu_\nu \bar{F}_{g^{\tau\tau} r^\mu_\nu} \;, \quad \mu_2(y) = \bar{F}_{{}^{(3)}\!R{}^{(3)}\!R} - \frac{1}{3} \bar{F}_{r^2} - \frac{2}{3} h^\mu_\nu \bar{F}_{{}^{(3)}\!R r^\mu_\nu} \;, \nonumber \\
\mu_3(y) &= \bar{F}_{r^2} \;, \quad \lambda_1(y)^\mu_\nu = 2 \bar{F}_{g^{\tau\tau} \sigma^\nu_\mu} \;, \quad \lambda_2(y)^\mu_\nu = 2 \bar{F}_{g^{\tau\tau} r^\nu_\mu} \;,\quad \lambda_3(y)^\mu_\nu = 2 \bar{F}_{K \sigma^\nu_\mu}\;, \nonumber \\ 
\lambda_4(y)^\mu_\nu &= 2 \bar{F}_{K r^\nu_\mu}  \;, \quad \lambda_5(y)^\mu_\nu = 2 \bar{F}_{{}^{(3)}\!R \sigma^\nu_\mu} \;, \quad \lambda_6(y)^\mu_\nu = 2 \bar{F}_{{}^{(3)}\!R r^\nu_\mu} \;. \nonumber 
\end{align}
From the estimates (\ref{eq:scaling1})--(\ref{eq:scaling3}) the EFT parameters defined above scale as
\begin{align}
f &\sim \mathcal{O}(1) \;, \quad \Lambda \sim  \mathcal{O}(M_\star^2 E^2) \;, \quad \beta \sim \alpha \sim \mathcal{O}(M_\star^2 E) \;, \quad \gamma \sim \mathcal{O}(M_\star^2) \;, \\
m_2^4 &\sim \mathcal{O}(M_\star^2 E^2) \;, \quad M_1^3 \sim \mathcal{O}(M_\star^2 E) \;, \quad M_3^2 \sim \mathcal{O}(M_\star^2) \;, \quad \mu_1^2 \sim \mathcal{O}(M_\star^2) \;,
\end{align}
and 
\begin{align}
M_4 \sim M_5 \sim \mathcal{O}(M_\star^2 E \Lambda_\star^{-2}) \;, \quad \mu_2 \sim \mu_3 \sim \mathcal{O}(M_\star^2 \Lambda_\star^{-2}) \;.
\end{align}
Besides, as we will see explicitly in Section~\ref{sec:bg_tadpole} the parameter $c$ scales as $c \sim \mathcal{O}(M_\star^2 E^2)$.

Written in terms of the parameters defined in (\ref{eq:EFT_parameters}) the consistency relations (\ref{eq:con2})--(\ref{eq:con9}) become
\begin{align}
\partial_i\Lambda + \bar{g}^{\tau\tau}\partial_i c - \frac{1}{2}M^2_\star {}^{(3)}\!\bar{R} \partial_i f + \frac{1}{3}\bar{K}(M^2_\star \bar{K}\partial_i f + 3\partial_i\beta) - \frac{1}{2} \bar{\sigma}^\mu_\nu (M^2_\star \bar{\sigma}^\nu_\mu \partial_i f & - 2\partial_i\alpha^\nu_\mu) \nonumber \\
&\hspace{4mm}+ \bar{r}^\mu_\nu\partial_i\gamma^{\nu}_\mu \simeq 0 \;, \label{eq:con_EFT1} \\
\partial_i c + m_2^4 \partial_i \bar{g}^{\tau\tau} + \frac{1}{2}(M_1^3 + \frac{1}{3} h^\mu_\nu \lambda_{1\mu}^\nu) \partial_i\bar{K} + \frac{1}{2}\lambda_{1\mu}^\nu \partial_i\bar{\sigma}^\mu_\nu  + (\mu_1^2 + \frac{1}{3}h^\mu_\nu \lambda_{2\mu}^\nu )\partial_i{}^{(3)}\!\bar{R} &+ \frac{1}{2}\lambda_{2\mu}^\nu \partial_i\bar{r}^\mu_\nu \simeq 0 \;, 
\end{align}
and
\begin{align}
\partial_i\beta + \frac{1}{2}(M_1^3 + \frac{1}{3}h^\mu_\nu \lambda^\nu_{1\mu})\partial_i \bar{g}^{\tau\tau} + \frac{2}{3}M_\star^2 \bar{K} \partial_i f + (M_2^2 &+ \frac{1}{3} M_3^2 + \frac{1}{3} h^\mu_\nu \lambda_{3 \mu}^\nu) \partial_i \bar{K} + \frac{1}{2} \lambda_{2 \mu}^\nu \partial_i\bar{\sigma}_\nu^\mu \nonumber \\ & \hspace{3mm} + \frac{1}{2} (M_4 + \frac{1}{3}M_5)\partial_i{}^{(3)}\!\bar{R} + \frac{1}{2}\lambda_{4 \mu}^\nu \partial_i\bar{r}^\mu_\nu \hspace{0.2mm} \simeq 0 \;, \\
\partial_i\alpha^\mu_\nu - M_\star^2 \bar{\sigma}^\mu_\nu \partial_i f + M_3^2 \partial_i\bar{\sigma}^\mu_\nu + \frac{1}{2} \lambda_{1 \nu}^\mu \partial_i\bar{g}^{\tau\tau} &+ \frac{1}{2} \lambda_{3 \nu}^\mu \partial_i\bar{K} + \frac{1}{2} \lambda_{5 \nu}^\mu \partial_i{}^{(3)}\!\bar{R} + \frac{1}{2}M_5 \partial_i\bar{r}^\mu_\nu \simeq 0 \label{eq:con_EFT2} \;, 
\end{align}
where $\partial_i$ denotes the partial derivative with respect to $x^i$. As we have explained before, these consistency relations guarantee that the EFT is invariant under the 3d diffeomorphism. Moreover, the relations above trivially hold in the case of a homogeneous background e.g.~the EFT of inflation/DE. We will check in Section~\ref{sec:Dictionary} that these consistency relations are automatically satisfied for Horndeski theories on any backgrounds. 

\subsection{Stueckelberg procedure}
In the last Subsection we have formulated a consistent EFT action in the unitary gauge on an inhomogeneous background. Despite the fact that each term of the EFT action can depend on the spatial coordinates $\vec{x}$, the consistency relations imposed on the EFT parameters guarantee that the residual gauge symmetry is the 3d diffeomorphism invariance. In this Section we will complete the formalism by discussing the Stueckelberg procedure~\footnote{This procedure will be useful for studying perturbations especially for even sector, but we leave this to the future work.}. 

It is well-known in the context of the EFT of inflation/DE \cite{Cheung:2007st,Gubitosi:2012hu} that one can use the Stueckelberg trick to restore the full 4d diffeomorphism invariance of the EFT action via introducing a Goldstone boson field $\pi$ that non-linearly realizes the time diffeomorphism symmetry (see \cite{Cusin:2017mzw} for the expressions up to second order).  Here we apply the same procedure introducing a Stueckelberg field $\pi$ via a transformation, $\tau \rightarrow \tau + \pi(\tau,\vec{x})$. Note that the spatial coordinates $\vec{x}$ remains unchanged, so that one does not have to worry about the spatially-dependent terms in our EFT action. 

Performing such a transformation in $\tau$ we find, for instance, that an arbitrary function of $\tau$ and $\vec{x}$, $f(\tau,\vec{x})$, can be replaced up to second order with
\begin{align}
f(\tau,\vec{x}) \rightarrow f(\tau,\vec{x}) + \dot{f}(\tau,\vec{x}) \pi + \frac{1}{2} \ddot{f}(\tau,\vec{x}) \pi^2 + \mathcal{O}(3) \;,
\end{align}
where an overdot denotes the partial derivative with respect to $\tau$, and the inverse metric with 
\begin{align}\label{eq:metric_stue}
g^{\mu\nu} \rightarrow (\delta^\mu_\alpha + \delta^\mu_\tau \partial_\alpha \pi)(\delta^\nu_\beta + \delta^\nu_\tau \partial_\beta \pi) g^{\alpha \beta} \;. 
\end{align}
In particular, the $(\tau\tau)$-component of the inverse metric transforms as 
\begin{align}\label{eq:stu_g_tautau}
g^{\tau\tau} \rightarrow g^{\tau\tau} + 2 g^{\tau \mu} \partial_\mu \pi + g^{\mu\nu} \partial_\mu \pi \partial_\nu \pi \;.
\end{align}
Using (\ref{eq:metric_stue}) one finds that the ADM variables defined in (\ref{eq:ADM_metric}) transform as
\begin{align}
N &\rightarrow N\bigg(1 - \dot{\pi} + N^i\partial_i \pi + \dot{\pi}^2 + \frac{1}{2} N^2 h^{ij}\partial_i\pi \partial_j \pi\bigg) + \mathcal{O}(3)  \;, \\
N^i &\rightarrow N^i(1 - \dot{\pi}) + N^2 h^{ij} (1 - 2\dot{\pi}) \partial_j \pi + \mathcal{O}(3) \;, \\
h_{ij} &\rightarrow h_{ij} - N_i\partial_j\pi - N_j \partial_i \pi - N^2 \partial_i\pi \partial_j \pi + \mathcal{O}(3) \;. 
\end{align}
Notice that in our case since the background we are going to consider is diagonal, the shift vector $N^i$ starts at first order in perturbations. It is also useful to note that both $\partial_\tau$ and $\partial_i$ transform as
\begin{align}
\partial_\tau \rightarrow (1 - \dot{\pi} + \dot{\pi}^2) \partial_\tau + \mathcal{O}(3) \;, \quad \partial_i \rightarrow  \partial_i - (1 - \dot{\pi}) \partial_i\pi \partial_\tau + \mathcal{O}(3) \;.
\end{align} 
Furthermore, the spatial Christoffel symbol,
\begin{align}
\Gamma_{ij}^\ell = \frac{1}{2}h^{\ell k}(\partial_i h_{k j} + \partial_j h_{k i} - \partial_k h_{i j}) \;,
\end{align}
transforms, up to second order, as
\begin{align}
\Gamma_{ij}^\ell &\rightarrow \Gamma_{ij}^\ell + \frac{1}{2} (N^\ell h^{mk} \partial_k \pi + N^m h^{\ell k} \partial_k \pi + N^2 h^{\ell k} h^{m n} \partial_k \pi \partial_n \pi) a_{ijm} - (1 - \dot{\pi}) h^{\ell k} \dot{h}_{k(i} \partial_{j)} \pi \nonumber \\ &- \frac{1}{2} [h^{\ell m} b_{ijm} + 2 N^\ell \partial_i \partial_j \pi + 2 \partial^\ell \pi \partial_{(i} N_{j)} - \partial^\ell N_j \partial_i \pi + 2 N^2 \partial^\ell \pi \partial_i \partial_j \pi + 2N h^{\ell m} c_{ijm}] + \mathcal{O}(3)\;, 
\end{align}
where we have defined 
\begin{align}
a_{ijm} &\equiv \partial_i h_{mj} + \partial_j h_{mi} - \partial_m h_{ij} \;, \\
b_{ijm} &\equiv \partial_i N_m \partial_j \pi + \partial_j N_m \partial_i \pi - \partial_m N_i \partial_j \pi \;, \\
c_{ijm} &\equiv \partial_m\pi \partial_i N \partial_j \pi  + \partial_m\pi \partial_j N \partial_i \pi - \partial_i\pi \partial_m N \partial_j \pi \;,
\end{align}
and the symmetrization of indices is represented by parentheses $()$. Notice that if the background value of $N$ does not depend on the spatial coordinates e.g.~the FLRW metric, $c_{ijm}$ then starts at third order in perturbations and second order otherwise. Finally, using the relations above we find the transformations of $K$, $K_{ij}$, ${}^{(3)}\!R$ and ${}^{(3)}\!R_{ij}$ from their definitions, 
\begin{align}
\delta K \rightarrow & \ \delta K - \dot{\bar{K}} \pi - \frac{1}{2} \ddot{\bar{K}} \pi^2 - (1 - \dot{\pi})h^{ij} N\partial_i\partial_j\pi + N(1 - \dot{\pi})h^{ij} \Gamma^\ell_{i j} \partial_\ell\pi  - \frac{5}{6}N^2 \bar{K}h^{ij}\partial_i\pi \partial_j \pi  \nonumber \\ &+ 2N h^{i j} \partial_{i} \dot{\pi} \partial_{j}\pi - 2(1 - \dot{\pi})h^{ij} \partial_{i} N \partial_{j}\pi + \dot{N} h^{i j} \partial_i \pi \partial_j \pi \nonumber \\ &- \frac{1}{2N} N^\ell h^{ij}  \dot{h}_{ij} \partial_\ell \pi + \mathcal{O}(3) \;, \\
\delta K_{\ j}^i \rightarrow  & \ \delta K^i_{\ j} - \dot{\bar{K}}^i_{\ j} \pi - \frac{1}{2} \dot{\bar{K}}^i_{\ j} \pi^2 - (1 - \dot{\pi})h^{i\ell} N\partial_\ell\partial_j\pi + N(1 - \dot{\pi})h^{i\ell} \Gamma^k_{\ell j} \partial_k\pi  \nonumber \\ & - \frac{1}{2}N^2 \bar{K}^i_{\ j} h^{k\ell}\partial_k\pi \partial_\ell \pi  - N^2 \bar{K}^{i\ell} \partial_\ell \pi \partial_j\pi + 2N h^{i \ell} \partial_{(\ell} \dot{\pi} \partial_{j)}\pi - 2(1 - \dot{\pi})h^{i\ell} \partial_{(\ell} N \partial_{j)}\pi \nonumber \\ &+ \dot{N} h^{i \ell} \partial_\ell \pi \partial_j \pi + \frac{1}{2N} (h^{\ell k} N^i - h^{i\ell} N^k) \dot{h}_{\ell j} \partial_k \pi - \frac{1}{2N} h^{i\ell} N^k \dot{h}_{k\ell} \partial_j\pi + \mathcal{O}(3) \;, \\
\delta {}^{(3)}\!R \rightarrow  & \ \delta {}^{(3)}\!R + \frac{1}{2}h^{\ell k}\dot{h}_{\ell k}h^{ij} \partial_i \partial_j \pi - \frac{1}{2}h^{\ell k}h^{ij}\dot{h}_{k i} \partial_\ell \partial_j \pi  + \mathcal{O}(2) \;, \\
\delta {}^{(3)}\!R_{ij} \rightarrow  & \ \delta {}^{(3)}\!R_{ij} + \frac{1}{2}h^{\ell k}\dot{h}_{\ell k} \partial_i \partial_j \pi + \frac{1}{2}h^{\ell k}(\dot{h}_{k i}\partial_\ell \partial_j \pi - \dot{h}_{kj} \partial_\ell \partial_i \pi - \dot{h}_{k i} \partial_\ell \partial_j \pi ) + \mathcal{O}(2) \;,
\end{align}
where in the expressions for the 3d curvature we have focused only on the terms that contain second derivatives acting on $\pi$, up to first order in perturbations. Note that from the relations above it is straightforward to obtain the transformations of the tensors $\sigma_{ij}$ and $r_{ij}$. We now comment on the formulas above as follows. First we see that the dependence on the spatial coordinates $\vec{x} = \{x^i\}$ is present through the background values of $K_{ij}$, ${}^{(3)}\!R_{ij}$ and their traces. For $K_{ij} \propto h_{ij}$ e.g.~a flat FLRW background the result coincides with the ones of \cite{Cusin:2017mzw}. As we emphasized before, the results we obtained in this Subsection can be applied to any inhomogeneous background metrics.

In the next Subsection we will use the Stueckelberg trick derived above and obtain the decoupling limit action i.e.~we neglect the mixing of scalar and metric perturbations. The resulting action will be useful to determine an energy scale at which, for example, the cubic self-interaction of $\pi$ becomes important. Note that the decoupling limit action is a good approximation on sufficiently small scales \cite{Motohashi:2019ymr,Cheung:2007st}.

\subsection{Decoupling limit action}
In this Subsection we focus on the decoupling limit action in which the scalar perturbation $\pi$ is decoupled from the metric perturbations. This situation happens when the scale of interest for perturbations is sufficiently shorter than the scale associated to the background. Therefore, the relevant terms in the EFT (\ref{eq:EFT}) to the quadratic and cubic Lagrangians of $\pi$ are given by
\begin{align}\label{eq:EFT_decoupling}
S = \int d^4x \sqrt{-g} \bigg[&\frac{M_\star^2}{2}f(y)R - \Lambda(y) - c(y)g^{\tau\tau} + \frac{1}{2} m_2^4(y) (\delta g^{\tau\tau})^2 + \frac{1}{3!} m_3^4(y) (\delta g^{\tau\tau})^3  \bigg] \;.
\end{align}
Here for simplicity we disregard those higher derivative terms involving $\delta K$ and $\delta {}^{(3)}\!R$. 
Note that including such terms, for instance $\delta g^{\tau\tau} \delta K$, the scale at which the non-linear self-interaction of $\pi$ becomes relevant is expected to be of the order of the cut-off scale of the EFT, $\Lambda_\star$.  

From the action (\ref{eq:EFT_decoupling}) one obtains the quadratic and cubic Lagrangian for $\pi$ using the Stueckelberg trick for $\delta g^{\tau\tau}$ (see eq.(\ref{eq:stu_g_tautau})), 
\begin{align}\label{eq:decoupling_pi}
S_{\pi} = \int d^4x \sqrt{-g} \bigg[(c + 2m_2^4)\dot{\pi}^2 - c h^{ij}\partial_i \pi \partial_j \pi + \bigg(2m_2^4 + \frac{8}{3} m_3^4\bigg)\dot{\pi}^3 - 2m_2^4 \dot{\pi} h^{ij}\partial_i \pi \partial_j \pi \bigg] \;,
\end{align}
where we have set $\bar{g}^{\tau\tau} = -1$, $\bar{g}^{\tau i} = 0$ and $\bar{g}^{ij} = h^{ij}$. Also, in (\ref{eq:decoupling_pi}) we have neglected the terms without derivatives acting on $\pi$ since those terms will be proportional to some positive powers of the energy scale $E$, eq.(\ref{eq:energy_E}), which we assume to be less than the energy scale associated to the $\pi$ field (for the validity of the decoupling limit). Introducing the sound speed, $c_s^2$,
\begin{align}\label{eq:c_s_decoupling}
\frac{1}{c_s^2} \equiv 1 + \frac{2m_2^4}{c} \;,
\end{align}
the action (\ref{eq:decoupling_pi}) becomes 
\begin{align}
S_\pi = \int d^4x \sqrt{-g} \bigg[\frac{c}{c_s^2}\bigg(\dot{\pi}^2 - c_s^2 h^{ij}\partial_i \pi \partial_j \pi \bigg) + c  \bigg(\frac{1}{c_s^2} - 1 \bigg) \dot{\pi} (\dot{\pi}^2 - h^{ij} \partial_i \pi \partial_j \pi) + \frac{8}{3} m_3^4 \dot{\pi}^3 \bigg] \;.
\end{align}
Notice that in this case where we assume that $c_s^2$ is not close to zero, one can safely neglect the terms coming from $\delta K$ or $\delta {}^{(3)}\!R$ since they are further suppressed by the cut-off scale.

In order to estimate the strong coupling scale of the action above, let us assume further that within a time scale associated to an energy scale of interest $E_\star$ the sound speed is a constant and $0 < c_s^2 < 1$. Given that one can rescale the spatial coordinates as $\vec{x} = c_s \vec{\tilde{x}}$, we obtain
\begin{align}
S_\pi = \int d\tau d^3\tilde{x} \sqrt{-g}~c_sc\bigg[\bigg(\dot{\pi}^2 - h^{ij}\tilde{\partial}_i \pi \tilde{\partial}_j \pi \bigg) + c_s^2  \bigg(\frac{1}{c_s^2} - 1 \bigg) \dot{\pi} (\dot{\pi}^2 - h^{ij} \tilde{\partial}_i \pi \tilde{\partial}_j \pi) + \frac{8}{3} \frac{m_3^4c_s^2}{c} \dot{\pi}^3 \bigg] \;. \label{eq:decoupling_rescale_pi}
\end{align}
We see that the theory at hand becomes strongly-coupled when the cubic terms are comparable to the quadratic terms. This happens when 
\begin{align}
\frac{\mathcal{L}_2}{\mathcal{L}_3} \sim 1 \quad \Rightarrow \quad |\dot{\pi}|\bigg|_{E_\star = E_{\rm Cubic}} \sim \frac{1}{{\rm max}\{c_s^2 m_3^4/c, 1-c_s^2\}} \;, \label{eq:L2-L3}
\end{align}
with $E_{\rm Cubic}$ being the strong coupling scale for the cubic interaction. Using the fact that for a given energy scale $E_\star$ the amplitude of fluctuations $\pi$ scales as 
\begin{align}
|\dot{\pi}| \sim \frac{E^2_\star}{\sqrt{c_s c}} \;,
\end{align}
and therefore, $E_{\rm Cubic}$ is approximately given by
\begin{align}\label{eq:E_cubic}
E_{\rm Cubic} \sim \frac{(c_s c)^{1/4}}{\sqrt{1 - c_s^2}} \sim \frac{(c_s M_\star^2 E^2)^{1/4}}{\sqrt{1 - c_s^2}} \;,
\end{align}
where in (\ref{eq:L2-L3}) we have taken $m_3^4/c \sim \mathcal{O}(1)$ and $c \sim \mathcal{O}(M_\star^2 E^2)$. This is indeed the energy scale up to which the quadratic terms in (\ref{eq:decoupling_rescale_pi}) remain to be the dominant terms in the Lagrangian. Clearly, the strong coupling scale can be lower than or equal to $E_{\rm Cubic}$ due to the presence of the higher-order derivative terms which may or may not be comparable to $\mathcal{L}_2$ below $E_{\rm Cubic}$~(\footnote{The typical example is the cubic Galilean operator or $\delta g^{\tau\tau} \delta K$ in the EFT action. This operator typically  becomes strongly coupled at energy scales parametrically lower than $E_{\rm Cubic} \sim (M_\star E)^{1/2}$, taking $c_s/(1 - c_s^2)^2 \sim 1$.}). It is worth pointing out that including the higher-order terms such as Cubic and Quartic Galilean operators one typically expects that in the non-linear regime the $\pi$-field becomes pathological in the presence of gravitational wave background, see \cite{Creminelli:2019kjy}.

Moreover, in the limit $c_s^2 \ll 1$ (\ref{eq:E_cubic}) seems to suggest that the system becomes infinitely strongly-coupled.  However, this is not the case since in such a limit of $c_s^2$ the terms involving $\delta K$ and $\delta {}^{(3)}\!R$ cannot be neglected and one needs to take these into account so that the strong coupling problem can be avoided. Without going into the detail, following the argument of \cite{Motohashi:2019ymr} one would conclude that in the limit $c_s^2 \ll 1$ the strong coupling scale can be as high as the scale $\mu$ (the scale associated to $\bar{\Phi}$). Specifically, the dispersion relation of $\pi$ taking into account, for example, the $\delta K^2$ term contains a term which goes as $k^4/\mu^2$~(\footnote{In the context of stealth solutions, i.e.~the background metric is the one of GR but the scalar profile is non-trivial, this $k^4$ contribution is called scordatura term.}), where $k$ is a spatial momentum of $\pi$. This additional $k^4$ term indeed evades the strong coupling problem we mentioned before and the perturbations are weakly coupled all the way up to the scale $\mu$~\cite{Arkani-Hamed:2003pdi,Motohashi:2019ymr}. 
\section{Dictionary}\label{sec:Dictionary}
In the previous Section we have written down the EFT action for perturbations in the unitary gauge and obtained the consistency relations under which the 3d diffeomorphism is unbroken. Here, we provide a connection between our EFT parameters and some covariant theories. For simplicity, we consider the Horndeski theories \cite{Horndeski:1974wa,Deffayet:2009mn} which are defined by four arbitrary functions of $\Phi$ and $X$,
\begin{align}\label{eq:Horndeski}
S = \int d^4x \sqrt{-g}~L_I \;,
\end{align}
where
\begin{align}
L_2 =~&G_2(\Phi,X) \;, \\
L_3 =~&G_3(\Phi,X) \Box\Phi \;, \label{eq:cubic_Horn} \\
L_4 =~&G_4(\Phi,X)\tilde{R} - 2 G_{4X}(\Phi,X) (\Box\Phi^2 - \nabla_\nu\nabla_\mu\Phi \nabla^\nu\nabla^\mu\Phi) \;, \label{eq:L_4_Horna} \\
L_5 =~&G_5(\Phi,X) \tilde{G}_{\mu\nu} \nabla^\nu\nabla^\mu\Phi \nonumber \\ &+ \frac{1}{3}G_{5X}(\Phi,X)(\Box \Phi^3 - 3 \Box\Phi \nabla_\nu\nabla_\mu\Phi \nabla^\nu\nabla^\mu\Phi + 2 \nabla_\nu\nabla_\mu\Phi \nabla^\sigma \nabla^\mu \Phi \nabla_\sigma \nabla^\nu \Phi) \;, \label{eq:L_5}
\end{align}
where $\tilde{G}_{\mu\nu} \equiv \tilde{R}_{\mu\nu} - \tilde{R}g_{\mu\nu}/2$ is the 4d Einstein tensor, $\nabla_\mu$ is the 4d covariant derivative as defined before and $\Box \equiv \nabla_\mu\nabla^\mu$. Here we define $G_{iX} \equiv \partial G_i/\partial X$ and $G_{i\Phi} \equiv \partial G_i/\partial \Phi$ with $i = 2,3,4,5$. To find the mapping between the Horndeski functions and those of the EFT action (\ref{eq:EFT}) one simply expresses those covariant Lagrangian above in terms of the building blocks of the EFT in the unitary gauge.  Note that here we do not assume any particular form of $\bar{g}^{\tau\tau}(\tau,\vec{x})$, therefore $\bar{X}$ can depend on the spatial coordinates. 

\subsection{$L_2$}
We first consider the function $G_2(\Phi,X)$. It is straightforward to find the Taylor coefficients defined in (\ref{eq:EFT_taylor}), 
\begin{align}
\bar{F} &= \bar{G}_2 \;, \\
\bar{F}_{g^{\tau\tau}} &= \dot{\bar{\Phi}}^2 \bar{G}_{2X}  \;, \\
\bar{F}_{g^{\tau\tau}g^{\tau\tau}} &= \dot{\bar{\Phi}}^4 \bar{G}_{2XX} \;. 
\end{align}
Using (\ref{eq:EFT_parameters}) we easily find
\begin{align}
\Lambda &= -\bar{G}_2 + \bar{g}^{\tau\tau} \dot{\bar{\Phi}}^2 \bar{G}_{2X} \;, \\
c &= - \dot{\bar{\Phi}}^2 \bar{G}_{2 X} \;, \\
m_2^4 &= \dot{\bar{\Phi}}^4 \bar{G}_{2XX} \;. 
\end{align}
From the expressions above, we see that the function $G_2(\Phi,X)$ only generates a polynomial of $\delta g^{\tau\tau}$ with spatially-dependent coefficients through $\bar{g}^{\tau\tau}(\tau,\vec{x})$. Note that in the case where $\bar{g}^{\tau\tau} = -1$ all the corresponding EFT parameters of $L_2$ become independent of $\vec{x}$.

\subsection{$L_3$}
Here we consider the cubic Horndeski operator, eq.(\ref{eq:cubic_Horn}). Following \cite{Gleyzes:2013ooa} let us introduce an auxiliary function $F_3(\Phi,X)$ via
\begin{align}
G_3(\Phi,X) = F_3(\Phi,X) + 2 X F_{3X}(\Phi,X) \;. 
\end{align}
The cubic action $S_3$ then becomes
\begin{align}
S_3 = \int d^4x \sqrt{-g}~\bigg[ F_3(\Phi,X) \Box \Phi  + 2 X F_{3X}(\Phi,X) \Box \Phi  \bigg]\;.
\end{align}
Integrating by parts for the first term and using the relation 
\begin{align}\label{eq:box_phi}
\Box \Phi = -\sqrt{-X} K + \frac{1}{2 \sqrt{-X}} n^\mu \partial_\mu X \;,
\end{align}
for the second term we thus obtain
\begin{align}
S_3 = \int d^4x \sqrt{-g}~\bigg[ -\bigg( F_{3X} \nabla_\mu X + F_{3\Phi} \nabla_\mu\Phi \bigg) \nabla^\mu \Phi  - 2 \sqrt{-X} X F_{3X} K + F_{3X} \nabla_\mu X \nabla^\mu\Phi  \bigg] \;.
\end{align}
Notice that the first term and the last term cancel each other.  Therefore, the action above is reduced to
\begin{align}\label{S_3_covariant}
S_3 = \int d^4x \sqrt{-g}~\bigg[ - F_{3\Phi} X + 2 (-X)^{3/2} F_{3X} K  \bigg] \;.
\end{align}
We see that now we have written the action in terms of the spatially covariant quantities and it is clearly invariant under 3d diffeomorphism. It is actually clear from (\ref{S_3_covariant}) that the action $S_3$ corresponds to the EFT operator with one factor of $\delta K$ together with a polynomial function of $\delta g^{\tau\tau}$, therefore the parameters such as $M_2$ and $M_3$ are expected to vanish. In particular, we determine the Taylor coefficients (\ref{eq:EFT_taylor}),
\begin{align}
\bar{F} &= -\bar{g}^{\tau\tau}\dot{\bar{\Phi}}^2 \bar{F}_{3 \Phi} + 2 (-\bar{g}^{\tau\tau})^{3/2} \dot{\bar{\Phi}}^3 \bar{K} \bar{F}_{3 X} \;, \label{L3_1} \\
\bar{F}_{g^{\tau\tau}} &=  -\sqrt{-\bar{g}^{\tau\tau}} \bar{K} \dot{\bar{\Phi}}^3 \bar{G}_{3X} - \dot{\bar{\Phi}}^2 \bar{G}_{3\Phi} + \bar{g}^{\tau\tau}\dot{\bar{\Phi}}^4 \bar{F}_{3X\Phi}  \label{L3_2} \;, \\
\bar{F}_K &= - 2 (-\bar{g}^{\tau\tau})^{3/2} \dot{\bar{\Phi}}^3 \bar{F}_{3X}  \;, \\
\bar{F}_{g^{\tau\tau}K} &= -\sqrt{-\bar{g}^{\tau\tau}} \dot{\bar{\Phi}}^3 \bar{G}_{3X}  \;, \\
\bar{F}_{g^{\tau\tau}g^{\tau\tau}} &= -\frac{1}{2} \dot{\bar{\Phi}}^4 \bar{G}_{3 \Phi X} + \frac{1}{2\sqrt{-\bar{g}^{\tau\tau}}}  \dot{\bar{\Phi}}^3 \bar{K} \bar{G}_{3 X} - \sqrt{-\bar{g}^{\tau\tau}} \dot{\bar{\Phi}}^5 \bar{K} \bar{G}_{3XX} + \frac{1}{2} \dot{\bar{\Phi}}^4 \bar{F}_{3\Phi X}  \;.
\end{align}
From the expressions above one finds the EFT parameters,
\begin{align}
\Lambda &= (-\bar{g}^{\tau\tau})^{3/2}\dot{\bar{\Phi}}^3 \bar{K} \bar{G}_{3X} - (-\bar{g}^{\tau\tau})^2 \dot{\bar{\Phi}}^4 \bar{F}_{3X\Phi} \;, \label{eq:Lambda_G3} \\
c &= \dot{\bar{\Phi}}^2 \bar{G}_{3\Phi} + \sqrt{-\bar{g}^{\tau\tau}} \dot{\bar{\Phi}}^3 \bar{K} \bar{G}_{3X} - \bar{g}^{\tau\tau} \dot{\bar{\Phi}}^4 \bar{F}_{3X\Phi}  \;, \\
\beta &= -2 (-\bar{g}^{\tau\tau})^{3/2} \dot{\bar{\Phi}}^3 \bar{F}_{3X}  \;, \\
M_1^3 &= -2 \sqrt{-\bar{g}^{\tau\tau}} \dot{\bar{\Phi}}^3 \bar{G}_{3X}  \;, \\
m_2^4 &= -\frac{1}{2} \dot{\bar{\Phi}}^4 \bar{G}_{3 \Phi X} + \frac{1}{2\sqrt{-\bar{g}^{\tau\tau}}}  \dot{\bar{\Phi}}^3 \bar{K} \bar{G}_{3 X} - \sqrt{-\bar{g}^{\tau\tau}} \dot{\bar{\Phi}}^5 \bar{K} \bar{G}_{3XX} + \frac{1}{2} \dot{\bar{\Phi}}^4 \bar{F}_{3\Phi X} \;. \label{eq:m_2_L3}
\end{align}
The EFT parameters associated with the terms higher-order in $\delta K$ are not present as expected. Note that the presence of the auxiliary function $F_3(\Phi,X)$ in the expressions above can be removed by using the relations (see Appendix A of \cite{Cusin:2017mzw})
\begin{align}\label{eq:F_3_G_3}
F_{3X} =  - \frac{1}{2 (-X)^{3/2}} \int dX' \sqrt{-X'} G_{3X'} \;, \quad F_{3\Phi} = -\frac{1}{\sqrt{-X}} \int dX' \frac{G_{3\Phi}}{2 \sqrt{-X'}} \;.
\end{align}
In addition, the fact that the function $F_3$ enters (\ref{eq:Lambda_G3})--(\ref{eq:m_2_L3}) with $X$-derivative or $\Phi$-derivative implies that the replacement (\ref{eq:F_3_G_3}) suffices to get rid of the function $F_3$ in (\ref{eq:Lambda_G3})--(\ref{eq:m_2_L3}).

It is interesting to notice that in (\ref{eq:m_2_L3}) there is a dependence on $\bar{K}$ which in general is a function of $\tau$ and $\vec{x}$. This means that the inhomogeneity of a background metric generally affects the sound speed $c_s^2$ (see eq.(\ref{eq:c_s_decoupling})) through $m_2^4$ and $c$ even under the assumption $\bar{X}=\bar{X}(\tau)$. 

\subsection{$L_4$}
Let us now consider $L_4$, eq.(\ref{eq:L_4_Horna}). It has been shown that $L_4$ can be rewritten in terms of spatially covariant quantities \cite{Gleyzes:2013ooa},
\begin{align}
L_4 = G_4 {}^{(3)}\!R + (2 X G_{4X} - G_4)\mathcal{K}_2 - 2 \sqrt{-X} G_{4 \Phi} K \label{eq:quartic} \;,
\end{align}
where $\mathcal{K}_2 \equiv K^2 - K_{\mu\nu} K^{\mu\nu} = 2K^2/3 - \sigma_{\mu\nu} \sigma^{\mu\nu}$. Note that the Lagrangians (\ref{eq:L_4_Horna}) and (\ref{eq:quartic}) are equivalent up to a boundary term due to an integration by parts. Although it is lengthy to find all the expressions for $L_4$, it is  straightforward to apply the same method done for $L_3$ to find the corresponding Taylor coefficients and the EFT parameters. Their explicit expressions can be found in Appendix~\ref{app:EFT_L4_L5}. They depend on the function $\bar{G}_4$ and its derivatives. 

As expected, within the quartic Horndeski operator one has $M_2^2 + M_3^3 = 0$ and no higher derivative terms i.e~$M_4 = M_5 = \mu_2 = \mu_3 = 0$. We will comment on the relation between $M_2^2$, $M_3^2$ and $\mu_1^2$ at the end of this Section together with $L_5$.

Similar to the $L_3$ case, the fact that $m_2^4$ depends on $\bar{K}$ implies that the sound speed of $\pi$ gets affected by the inhomogeneity of the black hole background even under the assumption $\bar{X}=\bar{X}(\tau)$. Moreover, the spatially-dependent background metric also generates the $\lambda_1$-operator in the EFT, which was not present in the EFT on the flat FLRW background. 

\subsection{$L_5$}
We now consider the quintic Lagrangian (\ref{eq:L_5}). Similar to $L_4$, $L_5$ can be rewritten as \cite{Gleyzes:2013ooa}
\begin{align}
L_5 = &-\sqrt{-X}~F_5 \mathcal{R}_3- \frac{1}{3} (-X)^{3/2} G_{5X} \mathcal{K}_3 + \frac{1}{2} X (G_{5\Phi} - F_{5\Phi}) {}^{(3)}\!R  + \frac{1}{2} X G_{5 \Phi} \mathcal{K}_2 \label{eq:quintic1} \;,
\end{align}
where an auxiliary function $F_5(\Phi,X)$ is defined by $G_{5X} \equiv F_{5X} + \frac{F_5}{2 X} $, $\mathcal{R}_3 \equiv K^{\mu\nu} {}^{(3)}\!R_{\mu\nu} - K {}^{(3)}\!R/2$
and $\mathcal{K}_3 \equiv K^3 - 3 K K_{\mu\nu} K^{\mu\nu} + 2 K_{\mu\nu} K^{\mu\alpha} K^\nu_{\alpha}$. Again, note that because of a few integration by parts the Lagrangians (\ref{eq:L_5}) and (\ref{eq:quintic1}) are equivalent up to boundary terms.

Notice that as pointed in \cite{Gleyzes:2013ooa} the last two terms on the RHS of (\ref{eq:quintic1}) can be recast in a form similar to the first two terms of (\ref{eq:quartic}) with the choice $F_4 = X(G_{5\Phi} - F_{5\Phi})/2$. The terms $\mathcal{R}_3$ and $\mathcal{K}_3$ in (\ref{eq:quintic1}) can be rewritten as
 \begin{align}
\mathcal{R}_3 &=  \sigma_{\mu\nu} r^{\mu\nu} - \frac{1}{6} K {}^{(3)}\!R \;, \\
\mathcal{K}_3 &= \frac{2}{9} K^3 - K \sigma_{\mu\nu}\sigma^{\mu\nu} + 2 \sigma_{\mu\nu} \sigma^{\mu\alpha} \sigma^\nu_\alpha \;.
 \end{align} 
Straightforwardly, we obtain all the Taylor coefficients and the corresponding EFT parameters of $L_5$, see Appendix~\ref{app:EFT_L4_L5} for explicit expressions. Note that similar to the case of $L_4$ one finds $M_2^2 + M_3^2 = 0$ and the $\lambda$'s terms are induced by non-vanishing values of $\bar{\sigma}_{\mu\nu}$ and $\bar{r}_{\mu\nu}$.  

\subsection{The resulting EFT}
Here we write down the EFT action of perturbations which corresponds to the Horndeski theories (\ref{eq:Horndeski}). Using the results of the previous Subsections one finds that the EFT action (\ref{eq:EFT}) is reduced to 
\begin{align}
S = \int d^4x \sqrt{-g} \bigg[&\frac{M^2_\star}{2}f(y)R - \Lambda(y) - c(y)g^{\tau\tau} - \beta(y) K  - \alpha^\mu_{\nu}(y)\sigma^{\nu}_\mu - \gamma^\mu_{\nu}(y)r^{\nu}_\mu \nonumber \\ 
& + \frac{1}{2} m_2^2(y) (\delta g^{\tau\tau})^2  + \frac{1}{2} M_1^3(y) \delta g^{\tau\tau} \delta K - \frac{1}{2} m_4^2(y) (\delta K^2 - \delta K^\mu_\nu \delta K^\nu_\mu)  \nonumber \\ & + \frac{1}{2}\mu_1^2(y) \delta g^{\tau\tau} \delta {}^{(3)}\!R  + \tilde{m}_4 (\delta K \delta {}^{(3)}\!R/2 
- \delta K^\mu_\nu \delta {}^{(3)}\!R^\nu_\mu) \nonumber \\ & + \frac{1}{2} \lambda_1(y)^\nu_\mu \delta g^{\tau\tau} \delta K^\mu_\nu +  \tilde{\lambda}(y)^\nu_\mu (\delta g^{\tau\tau} \delta {}^{(3)}\!R^\mu_\nu/2 -  \delta K \delta K^\mu_\nu /3)  \bigg] \;, \label{eq:EFT_Horn}
\end{align}
where $m_4^2 = -M_2^2 = M_3^2$, $\tilde{m}_4 = M_4 = -M_5/2$ and $\lambda_2(y)^\mu_\nu = -3 \lambda_3(y)^\mu_\nu/2 = \tilde{\lambda}^\mu_\nu$. The terms $\lambda_1(y)^\mu_\nu$ and $\tilde{\lambda}(y)^\mu_\nu$ survive because of the non-vanishing values of $\bar{\sigma}_{\mu\nu}$ and $\bar{r}_{\mu\nu}$, while they are not present in the EFT of inflation/DE. In addition, within the Horndeski theories one finds that $m_4^2 = 2(-\bar{g}^{\tau\tau}) \mu_1^2 - \sqrt{-\bar{g}^{\tau\tau}} \dot{\bar{\Phi}} \bar{K} \bar{F}_5 /3$~(\footnote{Using the integration by parts one is able to remove the extra term and the result coincides with the one found in \cite{Gleyzes:2013ooa} with $\bar{g}^{\tau\tau} = -1$.}).  

As a consistency check, we confirmed that the consistency relations (\ref{eq:con_EFT1})--(\ref{eq:con_EFT2}) are automatically satisfied, as it should be for any covariant theories such as Horndeski and DHOST theories. Therefore, our EFT action corresponding to the Horndeski theories is invariant under the 3d diffeomorphism as expected. It is worth noting that one can straightforwardly generalize our result of this Section to beyond Horndeski theories and DHOST theories.

This EFT action (\ref{eq:EFT_Horn}) can in principle be applied to a generic inhomogeneous background and in particular can be used to study black holes. We leave the analysis of black hole perturbations to future work~\cite{Mukohyama:2022skv}.

\section{Background equations for spherically symmetric, static black holes}\label{sec:bg_tadpole}
In this Section for simplicity we set $\gamma^\mu_\nu=0$ in (\ref{eq:EFT_Horn}) while keeping other EFT coefficients as general functions, which satisfy the consistency relations (\ref{eq:con_EFT1})--(\ref{eq:con_EFT2}) but are not necessarily given by the dictionary based on Horndeski theories. This assumption of vanishing $\gamma^\mu_\nu$ is motivated by the fact that in the context of Horndeski theories, the $Z_2$ symmetry, $\Phi \rightarrow -\Phi$, and the shift symmetry, $\Phi \rightarrow \Phi + const.$~\footnote{The shift symmetry guarantees that a background metric is time-independent. Without imposing the shift symmetry a fine-tuning among EFT parameters is needed to achieve a stationary black hole background with timelike scalar profile On the other hand, the $Z_2$ symmetry simplifies the analysis.}, forbid one to include the functions $G_3$ and $G_5$, and therefore force the $\gamma^\mu_\nu$ term in (\ref{eq:EFT_Horn}) to vanish. In principle, one could keep all the tadpole terms, but it would only lead to slight complications in the equations of motion.

With this setup, the terms in (\ref{eq:EFT_Horn}) which determine the background dynamics of the metric and the scalar field are
\begin{align}\label{eq:EFT_background_SRH}
S_{\rm tadpole} = \int d^4x \sqrt{-g} \bigg[\frac{M^2(y)}{2}R - \Lambda(y) - c(y)g^{\tau\tau} - \beta(y) K  - \alpha^\mu_{\nu}(y)\sigma^{\nu}_\mu \bigg] \;,
\end{align}
where $M^2(y) \equiv M^2_\star f(y)$. Before writing down the background equations of motion, let us discuss the last two terms on the RHS of (\ref{eq:EFT_background_SRH}). First, for the traceless background tensor $\alpha^{\mu}_{\nu}(y)$, it is convenient to assume that $\alpha^{\mu}_{\nu}(y) = \alpha(y) \bar{\sigma}^{\mu}_\nu$ with $\alpha(y)$ being an arbitrary function of $y = \{\tau, \rho\}$. It then follows from (\ref{eq:EFT_background_SRH}) that 
\begin{align}\label{eq:S_tadpole_final}
S_{\rm tadpole} = \int d^4x \sqrt{-g} \bigg[\frac{M^2(y)}{2}R - \Lambda(y) - c(y)g^{\tau\tau} - \tilde{\beta}(y) K - \alpha(y)\bar{K}^{\mu}_\nu K^\nu_{\mu}  \bigg] \;,
\end{align}
where $\tilde{\beta} \equiv \beta - \alpha\bar{K}/3$. For example, the quartic Horndeski function $G_4(X)$ gives $\tilde{\beta}(y) = -4\bar{X}\bar{G}_{4X} \bar{K}$. From now on, we consider the coefficients in this tadpole action as general functions, which satisfy the consistency relations (\ref{eq:con_EFT1})--(\ref{eq:con_EFT2}) with $\gamma^{\mu}_{\nu}=0$, $\alpha^{\mu}_{\nu}=\alpha\bar{\sigma}^{\mu}_{\nu}$ and $\beta=\tilde{\beta}+\alpha\bar{K}/3$ but are not necessarily given by the dictionary based on Horndeski theories. 

The Einstein equations of (\ref{eq:S_tadpole_final}) are then given by
\begin{align}\label{eq:eins}
(\bar{G}_{\mu\nu} - \bar{\nabla}_\mu \bar{\nabla}_\nu + \bar{g}_{\mu\nu}\bar{\Box}) M^2 - \bar{T}_{\mu\nu} = 0 \;,
\end{align}
where the background stress-energy tensor $\bar{T}_{\mu\nu}$ associated with the tadpole terms in (\ref{eq:S_tadpole_final}) is given by 
\begin{align}
\bar{T}_{\mu\nu} = -(c \bar{g}^{\tau\tau} + \Lambda \ + \ &\alpha \bar{K}^\alpha_{\beta}\bar{K}^{\beta}_\alpha) \bar{g}_{\mu\nu} + 2 c \delta_\mu^\tau \delta_\nu^\tau + 2 \alpha \bar{K}^\beta_{\mu} \bar{K}_{\beta \nu} - 2 \bar{n}_{(\beta}\partial_{\alpha)}\tilde{\beta} \delta^\beta_{\mu}\delta_\nu^\alpha + (\bar{g}_{\mu\nu} - \bar{n}_\mu \bar{n}_\nu) \bar{n}^\alpha \partial_\alpha \tilde{\beta}  \nonumber \\ 
 &+\alpha \bar{K}^\alpha_{\beta}\bar{K}^\beta_{\alpha} \bar{n}_\mu \bar{n}_\nu - \bar{\nabla}_\lambda (\alpha \bar{K}^\lambda_\mu \bar{n}_\nu) - \bar{\nabla}_\lambda(\alpha \bar{K}^\lambda_\nu \bar{n}_\mu) + \bar{\nabla}_\lambda (\alpha \bar{K}_{\mu\nu} \bar{n}^\lambda) \label{eq:T_munu_EFT} \;.
\end{align}

We now assume the background metric to be spherically symmetric and static in the sense of the usual Schwarzschild coordinates, namely, 
\begin{align}
ds^2 = -A(r) dt^2 + \frac{dr^2}{B(r)} + r^2 d\Omega^2 \;,
\end{align}
where $d\Omega^2 = d\theta^2 + \sin^2\theta d\phi^2$. The metric above can be rewritten in the Lema\^{\i}tre coordinates  \cite{Lemaitre:1933gd,Takahashi:2021bml}
\begin{align}\label{eq:metric_ansatz}
ds^2 = -d\tau^2 + [1 - A(r)] d\rho^2+ r^2 d\Omega^2 \;,
\end{align}
where the transformations are given by
\begin{align} \label{eq:taurho-tr}
d\tau = dt + \sqrt{\frac{1 - A(r)}{A(r)B(r)}}~dr \;, \quad d\rho = dt + \frac{dr}{\sqrt{A(r)B(r)(1 - A(r))}} \;.
\end{align}
From expressions above one obtains
\begin{align}
d(\rho - \tau) = \sqrt{\frac{A}{B(1 - A)}}~dr \;,
\end{align}
which clearly tells us that the original coordinate $r$ is a function of $\rho - \tau$, with 
\begin{align}
\partial_\rho r = -\dot{r} = \sqrt{\frac{B(1 - A)}{A}} \;.
\end{align}
We then set $\bar{\Phi}(\tau) = \mu^2 \tau$ so that $\bar{X}=-\mu^4$, where $\mu$ is a constant. A simple example for this type of background is shown in Appendix~\ref{app:simple_background}. Furthermore, the background quantities associated to the metric (\ref{eq:metric_ansatz}) such as $\bar{K}$ and ${}^{(3)}\!\bar{R}$ are given in Appendix~\ref{app:Background quantities}, considering $A$ and $B$ as arbitrary functions of $r$. The Einstein equations (\ref{eq:eins}) then determine relations between the functions in the background and the EFT coefficients.

We make one more assumption to simplify the analysis. In the context of Horndeski theories, the quintic Horndeski operator is absent if we assume the shift and $Z_2$ symmetries, and in this case the parameter $\alpha$ defined above for the background (\ref{eq:metric_ansatz}) depends only on $\tau$. We thus for simplicity restrict our considerations to the case where $\alpha = \alpha(\tau)$ which will simplify the analysis when one solves for the background equations of motion, 
\begin{align}\label{eq:S_tadpole_simple}
S_{\rm tadpole} = \int d^4x \sqrt{-g} \bigg[\frac{M^2(y)}{2}R - \Lambda(y) - c(y)g^{\tau\tau} - \tilde{\beta}(y)K - \alpha(\tau)\bar{K}^{\mu}_\nu K^\nu_{\mu}  \bigg] \;.
\end{align}
From eq.(\ref{eq:T_munu_EFT}) the non-vanishing components of $\bar{T}^\mu_{\ \nu}$ then read
\begin{align}
\bar{T}^\tau_{\ \tau} &= H_1(A,B;r) M^2 + H_2(A,B;r) (M^2)^\cdot + H_3(A,B;r) \partial_\rho(M^2) + \frac{\partial_\rho^2(M^2)}{1 - A} \;, \label{eq:T_tautau} \\
\bar{T}^\tau_{\ \rho} &= H_4(A,B;r) M^2 + H_5(A,B;r) \partial_\rho(M^2) + \partial_\rho(M^2)^{\cdot} \;, \label{eq:T_tau_rho} \\
\bar{T}^\rho_{\ \rho} &= H_6(A,B;r) M^2 + H_7(A,B;r)(M^2)^\cdot + H_8(A,B;r)\partial_\rho(M^2) - (M^2)^{\cdot\cdot}\;, \label{eq:T_rhorho} \\ 
\bar{T}^\theta_{\ \theta} &=   H_9(A,B;r) M^2 + H_{10}(A,B;r) (M^2)^\cdot + H_{11}(A,B;r) \partial_\rho(M^2) - 4 (M^2)^{\cdot\cdot} + \frac{4}{1 - A} \partial_\rho^2(M^2)\;, \label{eq:T_angular} \\
\bar{T}^\phi_{\ \phi} &= \bar{T}^\theta_{\ \theta}\;,\nonumber
\end{align}
where concrete expressions of the functions $H$'s are given in Appendix~\ref{app:functions_H}.

Substituting (\ref{eq:T_tautau})--(\ref{eq:T_angular}) into (\ref{eq:eins}) one obtains the equations determining $\Lambda(y)$ and $c(y)$,
\begin{align}
\Lambda(y) = \alpha E_1(A,B;r) - \frac{2 A'}{\sqrt{1 - A}}\sqrt{\frac{B}{A}}~\dot{\alpha} - \frac{\dot{\tilde{\beta}}}{4} - \frac{1}{4} (2H_1 - H_6) M^2 - \frac{1}{4} (2H_2 - H_7) (M^2)^\cdot \nonumber \\  - \frac{1}{4} (2H_3 - H_8) \partial_\rho(M^2)  - \frac{\partial_\rho^2(M^2)}{2(1 - A)} - \frac{1}{4} (M^2)^{\cdot\cdot} \;, \label{eq:tadpole_Lambda}
\end{align}
and
\begin{align}
c(y) = \alpha E_2(A,B;r) - \frac{ 4 A'}{4\sqrt{1 - A}}\sqrt{\frac{B}{A}}~\dot{\alpha} - \frac{\dot{\tilde{\beta}}}{2} - \frac{1}{2} (H_1 - H_6) M^2 - \frac{1}{2} (H_2 - H_7) (M^2)^\cdot \nonumber \\ - \frac{1}{2} (H_3 - H_8) \partial_\rho(M^2)  - \frac{\partial_\rho^2(M^2)}{2(1 - A)} - \frac{1}{2} (M^2)^{\cdot\cdot} \;, \label{eq:tadpole_c}
\end{align}
where $E_1$ and $E_2$ are defined by
\begin{align}
E_1 &\equiv  \frac{B}{A^2r^2} \bigg[5(1 - A)A +  A r A' - \frac{A}{8(1 - A)} r^2 (A')^2 - r^2 (A')^2 + 2 A r^2 A'' \bigg] + \frac{A' B'}{ A} \;, \\
E_2 &\equiv \frac{B}{A^2r^2} \bigg[7(1 - A)A + 2 A r A' - \frac{5A}{8(1 - A)} r^2 (A')^2 - 2 r^2 (A')^2 + 4 A r^2 A'' \bigg] + \frac{2 A' B'}{A} \;.
\end{align}
Note that $A' = dA/dr$ and $B' = dB/dr$. Eq.(\ref{eq:T_tau_rho}) gives rise to an equation for $\alpha$:
\begin{align}
\frac{A'}{2} \sqrt{\frac{B}{A(1 - A)}} \partial_\rho(M^2) + \frac{4}{A}\bigg[\frac{2B(1 - A)}{r^2} + \frac{B A'}{r} - \frac{B (A')^2 (1 - 2A)}{A (1 - A)} + A'B' + 2 B A'' \bigg] \alpha \nonumber \\ - \partial_\rho\tilde{\beta} - H_4 M^2 - \partial_\rho(M^2)^\cdot = 0 \;, \label{eq:alpha_EFT}
\end{align}
in which $\alpha$ can be algebraically solved in terms of $M^2$ and its derivatives, and $\partial_\rho\tilde{\beta}$. Furthermore, taking eq.(\ref{eq:T_rhorho}) subtracted with (\ref{eq:T_angular}) provides an equation:
\begin{align}
&\bigg[-\frac{16(1 - A)}{r^2} + \frac{3 A'^2}{1 - A} - 8 A'' + \frac{6 A'}{r} \bigg]\alpha  + \bigg[\frac{8 A' }{\sqrt{1 - A}} + \frac{16 \sqrt{1 - A}}{r} \bigg]\dot{\alpha}  - (H_6 - H_9) M^2  \nonumber \\ 
& - (H_7 - H_{10}) (M^2)^\cdot - (H_8 - H_{11}) \partial_\rho (M^2) - 3 (M^2)^{\cdot\cdot} + \frac{4}{1 - A} \partial^2_\rho (M^2) = 0 \;. \label{eq:beta_EFT}
\end{align}
Using the solution of $\alpha$ obtained from (\ref{eq:alpha_EFT}) in (\ref{eq:beta_EFT}) gives rise to a differential equation for $\tilde{\beta}$. We can then in principle solve for $\tilde{\beta}$ in terms of $M^2$ and its derivatives. We then plug this solution of $\tilde{\beta}$ together with the solution of $\alpha$ from (\ref{eq:alpha_EFT})  back into (\ref{eq:tadpole_Lambda}) and (\ref{eq:tadpole_c}) to obtain $\Lambda$ and $c$ in terms of $M^2$ and its derivatives. Therefore, we are left with only one free parameter $M^2$, while the other parameters in the tadpole action (\ref{eq:S_tadpole_simple}) are fixed once both the function $M^2$ and the background metric are specified. Finally, since we have assumed for simplicity that $\alpha$ depends only on $\tau$, one needs to impose the condition that the $\rho$-derivative of the solution of $\alpha$ obtained from (\ref{eq:alpha_EFT}) vanishes.

It is interesting to note that if one sets $\alpha = \partial_\rho \beta = 0$ and $M^2 = M_{\rm Pl}^2$ (the former can be realized e.g. in theories involving at most the cubic Horndeski and the latter can be achieved by a conformal transformation of the metric), eq.(\ref{eq:alpha_EFT}) gives rise to $A(r) = (const.) B(r)$. 

We leave a full analysis of odd and even-parity perturbations around the background (\ref{eq:metric_ansatz}) to future work \cite{Mukohyama:2022skv}.

\section{Conclusions} \label{sec:conclusions}
We formulated the Effective Field Theory (EFT) of black hole perturbations on an inhomogeneous background metric with a \textit{timelike} scalar profile in the context of scalar-tensor theories. The construction of the EFT in the unitary gauge was based on the fact that the background of the scalar field spontaneously breaks the time diffeomorphism ($\tau$-direction), whereas the spatial diffeomorphism serves as the residual gauge symmetry that leaves constant-$\Phi$ hypersurfaces invariant. Similar to the EFT of DE/Inflation, in Section~\ref{sec2:EFT_inhomo} we pointed out that the generic action in the unitary gauge is an spacetime integral of a scalar function of quantities covariant under 3d diffeomorphism; those quantities are, for instance, the 3d curvature, the extrinsic curvature and $g^{\tau\tau}$. On the contrary, their perturbations such as $\delta g^{\tau\tau}$ and $\delta K$ by themselves are not good building blocks of our EFT since they do not transform covariantly under 3d diffeomorphism (this is simply due to the spatial dependence in their background values). We found that a consistent way of constructing the EFT is to expand the unitary gauge action around the inhomogeneous background metric and to impose a set of consistency relations among the expansion coefficients, which guarantees  that the spatial diffeomorphism is preserved at any orders in perturbation. The non-trivial consistency relations are those associated with the spatial derivatives, whereas those with $\tau$-derivative do not result in consistency relations. We then obtained the EFT action of black hole perturbations in the unitary gauge with the EFT parameters satisfying the consistency relations. Indeed, this EFT action is generic in the sense that one can apply this action to a generic spatially-dependent background metric with a timelike scalar profile. In Section~\ref{sec:Dictionary} we reported a connection of our EFT parameters and the Horndeski functions. The resulting dictionary is in fact applicable to any background metric with $\bar{g}^{\tau\tau}$ being an arbitrary function of the spacetime coordinates. In Section~\ref{sec:bg_tadpole} we determined the background equations of motion on a static, spherically symmetric background metric with additional shift- and $Z_2$-symmetries of $\Phi$. 

There are several future directions we would like to explore. As we have mentioned before, the immediate application of the formalism developed in this paper is to study odd- and even-parity perturbations of a spherically symmetric black hole \cite{Mukohyama:2022skv}. It would be nice to use our EFT method to determine a spectrum of black hole quasi-normal modes for both odd- and even-parity sectors~\footnote{See \cite{Tomikawa:2021pca} for recent work on black hole quasi-normal modes with time dependent scalar hair in shift-symmetric scalar-tensor theories. There, they assumed the background metric to be static and spherically symmetric, and studied only for odd-parity perturbations.}, and to compute the tidal Love number of black holes~\footnote{It was found that the Love number of Schwarzschild and Kerr black holes in GR vanishes, see e.g.~\cite{Hui:2020xxx,Charalambous:2021mea}. However, it was found to be non-zero in some classes of modified gravity theories \cite{Cardoso:2018ptl,Bernard:2019yfz}.}. Also, we expect that the result of this analysis would lead to an interesting phenomena regarding the propagation of GWs, whether or not there is a significant change to the previous constraints on the EFT of DE, using the bound on $c_T^2$ at LIGO/Virgo scales, due to the presence of an inhomogeneity. Another future prospect is to apply our EFT to a more realistic background metric---a stationary and axisymmetric black hole---and to study the dynamics of perturbations. In this case, one would expect that the propagation speed may depend on directions in general, but we leave the analysis to future work. Note that while the EFT for perturbations of a slowly rotating black hole with a \textit{spacelike} scalar profile was recently studied in \cite{Hui:2021cpm}, we are interested in the case with a \textit{timelike} scalar profile.

\section*{Acknowledgements}
It is a pleasure to thank K.~Takahashi and G.~Tambalo for useful discussions, and M.~A.~Butt for carefully reading the manuscript. V.~Y. would like to thank the hospitality of the Abdus Salam International Centre for Theoretical Physics (ICTP) during the initial stage of this project. This work was supported by World Premier International Research Center Initiative (WPI), MEXT, Japan.
The work of S.~M. was supported by JSPS KAKENHI No.\ 17H02890, No.\ 17H06359.


\appendix

\section{Simple example of background}
\label{app:simple_background}

One example of a static, spherically symmetric background is the Schwarzschild solution with the gravitational radius $r_s$ in Lema\^{\i}tre coordinates \cite{Lemaitre:1933gd},
\begin{align}\label{eq:metric_Schwarzschild}
ds^2 = -d\tau^2 + [1 - A(r)] d\rho^2+ r^2 d\Omega^2 \;,
\end{align}
where $A(r) = 1 - r_s/r$ and $d\Omega^2 = d\theta^2 + \sin^2\theta d\phi^2$. The area radius $r$ is related to ($\tau,\rho$) coordinates via
\begin{align}
r(\tau,\rho) = r_s^{1/3} \bigg[\frac{3}{2}(\rho - \tau) \bigg]^{2/3} \;,
\end{align}
thus clearly the metric (\ref{eq:metric_Schwarzschild}) does not have a translational invariance in $\rho$. In the context of $\mathcal{P}(X)$ theory the metric (\ref{eq:metric_Schwarzschild}) is a vacuum solution to the Einstein equation assuming that $\bar{\Phi} = \mu^2 \tau$ and $\mathcal{P}'(\bar{X}) = \mathcal{P}(\bar{X}) = 0$, see \cite{Mukohyama:2005rw}. For this particular solution we have 
\begin{align}\label{eq:metric_Schwarzschild_gtautauKR}
\bar{g}^{\tau\tau} = -1 \;, \quad \bar{K}(\tau,\rho) = -\frac{3}{2 r(\tau,\rho)} \sqrt{\frac{r_s}{r(\tau,\rho)}}  \;,  \quad {}^{(3)}\!\bar{R} = 0 \;.
\end{align}
We see that even for this simplest background metric any naive scalar function made out of $\delta K$ is not invariant under 3d diffeomorphism due to an explicit $\rho$ dependence in $\bar{K}$. As we pointed out before, imposing the consistency relations (\ref{eq:con_EFT1})--(\ref{eq:con_EFT2}) ensures the invariance of the EFT under the spatial diffeomorphim. 

Note that in \cite{Kobayashi:2014eva} the stealth Schwarzschild-de Sitter solution was obtained in a shift- and reflection- symmetric subclass of the Horndeski theory. See \cite{Takahashi:2019oxz} for a similar solution in the shift- and reflection-symmetric subclass of quadratic DHOST theories.  For DHOST theories even without the reflection symmetry of $\Phi$ the exact GR solutions with a non-trivial scalar profile was found in \cite{Takahashi:2020hso} as long as a set of conditions on the coupling functions is satisfied.


\section{Background quantities}

\subsection{$\bar{K}_{\mu\nu}$ and ${}^{(3)}\!\bar{R}_{\mu\nu}$}\label{app:Background quantities}
Here we express all the background quantities of the ansatz (\ref{eq:metric_ansatz}) with (\ref{eq:taurho-tr}). First, the trace of the extrinsic curvature and the non-vanishing components of $\sigma_{\mu\nu}$ are given by
\begin{align}
\bar{K} &= -\frac{2}{r} \sqrt{\frac{B}{A(1 - A)}} \bigg(1 - A - \frac{r A'}{4}\bigg) \;, \\
\bar{\sigma}^\rho_{\ \rho} &= \frac{2}{3 r} \sqrt{\frac{B}{A(1 - A)}} \bigg(1 - A + \frac{r A'}{2} \bigg)  \;, \\
\bar{\sigma}^\theta_{\ \theta} &= \bar{\sigma}^\phi_{\ \phi}=  -\frac{1}{3 r} \sqrt{\frac{B}{A(1 - A)}} \bigg(1 - A + \frac{r A'}{2}\bigg)\;. 
\end{align}
Using the traceless condition, $\bar{\sigma}^\mu_{\ \mu} = 0$, one finds that there is only one independent component of $\bar{\sigma}^\mu_{\ \nu}$. It is clear that even at the background level the traceless parts of $\bar{K}_{\mu\nu}$ are generally non-zero and depend on $\rho$. Similarly, one can calculate the background values of ${}^{(3)}\!R$ and $r_{\mu\nu}$:
\begin{align}
{}^{(3)}\!\bar{R} &= \frac{2}{r^2}\bigg(1 - \frac{B}{A} + \frac{B r A'}{A^2} - \frac{r B'}{A} \bigg)  \;, \\
\bar{r}^\rho_{\ \rho} &= - \frac{2}{3r^2} \bigg(1 - \frac{B}{A} - \frac{r B A'}{2 A^2} + \frac{r B'}{2 A} \bigg) \;, \\ 
\bar{r}^\theta_{\ \theta} &= \bar{r}^\phi_{\ \phi} = \frac{1}{3r^2} \bigg(1 - \frac{B}{A} - \frac{r B A'}{2 A^2} + \frac{r B'}{2 A} \bigg) \;.
\end{align}
Moreover, the condition $\bar{r}^\mu_{\ \mu} = 0$ leads to the fact that we have only one independent component of $\bar{r}_{\mu\nu}$. Notice that when $A = B$ all the components of ${}^{(3)}\!\bar{R}_{\mu\nu}$ vanish. 

\subsection{The functions $H_i(A,B;r)$}\label{app:functions_H}
Here we give the expressions of the functions $H$'s defined in (\ref{eq:T_tautau})--(\ref{eq:T_angular}). We have 
\begin{align}
H_1 &=  -\frac{1}{r(1 - A)} \bigg[ 1 + \frac{A^2 (1 - B)}{1 - A} - \frac{B(1 - 2A)}{1 - A} - (1- A) \frac{r B'}{A} + (1 - A)^2 \frac{r B A'}{A^2}\bigg] \;, \nonumber \\
H_2 &= \frac{(4 -  4A - A')}{2 r \sqrt{1 - A}} \sqrt{\frac{B}{A}}  \;, \quad H_3 = \frac{(4 -  4A + A')}{2 r (1 - A)^{3/2}} \sqrt{\frac{B}{A}} \;,  \nonumber \\
H_4 &= \frac{(1 - A)(A'B - AB')}{A^2 r} \;, \quad H_5 = - \frac{A'}{2}\sqrt{\frac{B}{A(1 - A)}} \;, \\
H_6 &= -\frac{1}{r^2}\bigg[1 - B \bigg(1 + \frac{r A'}{A}\bigg) - (1 - A)\frac{r B'}{A}  \bigg] \;, \quad H_7 = \frac{2}{r} \sqrt{\frac{B(1 - A)}{A}} \;,  \nonumber \\
H_8 &= \frac{2}{r} \sqrt{\frac{B}{A(1 - A)}}  \;, \quad H_9 = \frac{B'}{2r} - \frac{B (A')^2}{4A^2} + \frac{A'B'}{4A} + \frac{A'B}{2 rA} + \frac{B A''}{2 A}  \;,   \nonumber \\  
\quad H_{10} &= \frac{2}{r} \sqrt{\frac{B}{A}} \frac{(2 - 2A - rA')}{\sqrt{1 - A}} \;, \quad
H_{11} = \frac{2}{r} \sqrt{\frac{B}{A}} \frac{(2 - 2A + rA')}{(1 - A)^{3/2}} \;. \nonumber
\end{align}

\section{EFT parameters for $L_4$ and $L_5$}\label{app:EFT_L4_L5}
In this Appendix we report all the Taylor coefficients and the EFT parameters of the covariant Lagrangians (\ref{eq:quartic}) and (\ref{eq:quintic1}).

\subsection{$L_4$}
From eq.(\ref{eq:quartic}) the Taylor coefficients are given by
\begin{align}
\bar{F} = \ & \bar{G}_4 {}^{(3)}\!\bar{R} + (2 \bar{g}^{\tau\tau} \dot{\bar{\Phi}}^2 \bar{G}_{4X} - \bar{G}_4 ) \mathcal{\bar{K}}_2 - 2 \sqrt{-\bar{g}^{\tau\tau}} \dot{\bar{\Phi}} \bar{K} \bar{G}_{4 \Phi} \;, \quad \bar{F}_{{}^{(3)}\!R} = \bar{G}_4\;, \nonumber \\ 
\bar{F}_{g^{\tau\tau}} = \ & \dot{\bar{\Phi}}^2 \bar{G}_{4X} {}^{(3)}\!\bar{R} + \mathcal{G}_4 \mathcal{\bar{K}}_2 + \frac{1}{\sqrt{-\bar{g}^{\tau\tau}}} \dot{\bar{\Phi}} \bar{K} \bar{G}_{4\Phi}  - 2 \sqrt{-\bar{g}^{\tau\tau}} \dot{\bar{\Phi}}^3 \bar{K} \bar{G}_{4 \Phi X} \;, \nonumber \\
\bar{F}_K = \ & \frac{4}{3} \bar{K} (2 \bar{g}^{\tau\tau} \dot{\bar{\Phi}}^2 \bar{G}_{4X} - \bar{G}_4) - 2 \sqrt{-\bar{g}^{\tau\tau}} \dot{\bar{\Phi}} \bar{G}_{4 \Phi} \;, \quad \bar{F}_{\sigma^\mu_\nu} = -2 \bar{\sigma}^\nu_\mu (2 \bar{g}^{\tau\tau} \dot{\bar{\Phi}}^2 \bar{G}_{4X} - \bar{G}_4 ) \;, \nonumber  \\
\bar{F}_{g^{\tau\tau}g^{\tau\tau}} =\ &\dot{\bar{\Phi}}^4 \bar{G}_{4XX} {}^{(3)}\!\bar{R} + (3 \dot{\bar{\Phi}}^4 \bar{G}_{4XX} + 2 \bar{g}^{\tau\tau} \dot{\bar{\Phi}}^6 \bar{G}_{4 XXX}) \mathcal{\bar{K}}_2  + \frac{1}{2 (-\bar{g}^{\tau\tau})^{3/2}} \dot{\bar{\Phi}} \bar{K} \bar{G}_{4\Phi} \\ 
&  + \frac{2}{\sqrt{-\bar{g}^{\tau\tau}}} \dot{\bar{\Phi}}^3 \bar{K} \bar{G}_{4 \Phi X} - 2 \sqrt{-\bar{g}^{\tau\tau}} \dot{\bar{\Phi}}^5 \bar{K} \bar{G}_{4 \Phi XX} \;, \nonumber \\ 
\bar{F}_{g^{\tau\tau} K} = \ &\frac{4}{3} \bar{K} \mathcal{G}_4 + \frac{1}{\sqrt{-\bar{g}^{\tau\tau}}} \dot{\bar{\Phi}} \bar{G}_{4\Phi} - 2 \sqrt{-\bar{g}^{\tau\tau}} \dot{\bar{\Phi}}^3 \bar{G}_{4 \Phi X} \;, \quad 
\bar{F}_{g^{\tau\tau} \sigma^\mu_\nu} = -2 \bar{\sigma}^\nu_\mu \mathcal{G}_4 \;, \nonumber \\
\bar{F}_{g^{\tau\tau} {}^{(3)}\!R} = \ & \dot{\bar{\Phi}}^2 \bar{G}_{4X} \;, \quad 
\bar{F}_{KK} =  \frac{4}{3} (2 \bar{g}^{\tau\tau} \dot{\bar{\Phi}}^2 \bar{G}_{4X} -  \bar{G}_4) \;, \quad
\bar{F}_{\sigma^2} = -2(2 \bar{g}^{\tau\tau} \dot{\bar{\Phi}}^2 \bar{G}_{4X} - \bar{G}_4) \;, \nonumber 
\end{align}
where $\mathcal{G}_4 \equiv \dot{\bar{\Phi}}^2 \bar{G}_{4X} + 2 \bar{g}^{\tau\tau} \dot{\bar{\Phi}}^4 \bar{G}_{4XX}$. The EFT parameters defined in (\ref{eq:EFT}) are then given by
\begin{align}
M_\star^2 f = \ &2 \bar{G}_4 \;, \nonumber \\
\Lambda = \ & \bar{g}^{\tau\tau} \dot{\bar{\Phi}}^2 \bar{G}_{4X} ({}^{(3)}\!\bar{R} + 3 \mathcal{\bar{K}}_2) +  2 (-\bar{g}^{\tau\tau})^2 \dot{\bar{\Phi}}^4 \bar{G}_{4XX}\mathcal{\bar{K}}_2 - \sqrt{-\bar{g}^{\tau\tau}} \dot{\bar{\Phi}} \bar{K} \bar{G}_{4\Phi} + 2 (-\bar{g}^{\tau\tau})^{3/2} \dot{\bar{\Phi}}^3 \bar{K} \bar{G}_{4\Phi X}  \;, \nonumber \\
c = \ & -\dot{\bar{\Phi}}^2 \bar{G}_{4X} {}^{(3)}\!\bar{R} - \mathcal{G}_4 \mathcal{\bar{K}}_2- \frac{1}{\sqrt{-\bar{g}^{\tau\tau}}} \dot{\bar{\Phi}} \bar{K} \bar{G}_{4\Phi} + 2 \sqrt{-\bar{g}^{\tau\tau}} \dot{\bar{\Phi}}^3 \bar{K} \bar{G}_{4 \Phi X} \;, \nonumber  \\
\beta = \ & -\frac{8}{3}\bar{g}^{\tau\tau} \dot{\bar{\Phi}}^2 \bar{K} \bar{G}_{4X}  + 2 \sqrt{-\bar{g}^{\tau\tau}} \dot{\bar{\Phi}} \bar{G}_{4\Phi} \;, \quad 
\alpha^\mu_\nu =  4 \bar{g}^{\tau\tau} \dot{\bar{\Phi}}^2 \bar{G}_{4X} \bar{\sigma}^\mu_\nu \;, \nonumber  \\
m_2^4 = \ &\dot{\bar{\Phi}}^4 \bar{G}_{4XX} {}^{(3)}\!\bar{R} + (3 \dot{\bar{\Phi}}^4 \bar{G}_{4XX} + 2 \bar{g}^{\tau\tau} \dot{\bar{\Phi}}^6 \bar{G}_{4 XXX})\mathcal{\bar{K}}_2 + \frac{1}{2 (-\bar{g}^{\tau\tau})^{3/2}} \dot{\bar{\Phi}} \bar{K} \bar{G}_{4\Phi}   \\ 
&+ \frac{2}{\sqrt{-\bar{g}^{\tau\tau}}} \dot{\bar{\Phi}}^3 \bar{K} \bar{G}_{4 \Phi X} - 2 \sqrt{-\bar{g}^{\tau\tau}} \dot{\bar{\Phi}}^5 \bar{K} \bar{G}_{4 \Phi XX} \;, \nonumber \\ 
M_1^3 = \ & \frac{8}{3} \bar{K} \mathcal{G}_4  + \frac{2}{\sqrt{-\bar{g}^{\tau\tau}}} \dot{\bar{\Phi}} \bar{G}_{4\Phi} - 4 \sqrt{-\bar{g}^{\tau\tau}} \dot{\bar{\Phi}}^3 \bar{G}_{4 \Phi X} \;, \quad
M_2^2 = 4 \bar{g}^{\tau\tau} \dot{\bar{\Phi}}^2 \bar{G}_{4X} \;, \nonumber \\
M_3^2 = \ & -4 \bar{g}^{\tau\tau} \dot{\bar{\Phi}}^2 \bar{G}_{4X} \;, \quad
\mu_1^2 = 2\dot{\bar{\Phi}}^2 \bar{G}_{4X} \;, \quad
\lambda^\mu_{1 \ \nu} = -4 \mathcal{G}_4 \bar{\sigma}^\mu_\nu \;. \nonumber
\end{align}

\subsection{$L_5$}
 Similar to the previous Subsection, the Taylor coefficients of (\ref{eq:quintic1}) are given by 
 \begin{align}
 \bar{F} = \ &-\sqrt{-\bar{g}^{\tau\tau}} \dot{\bar{\Phi}} \bar{F}_5 \mathcal{\bar{R}}_3 + \bar{F}_4 {}^{(3)}\!\bar{R} + \frac{1}{2} \bar{g}^{\tau\tau} \dot{\bar{\Phi}}^2 \bar{G}_{5\Phi} \mathcal{\bar{K}}_2  - \frac{1}{3} (-\bar{g}^{\tau\tau})^{3/2} \dot{\bar{\Phi}}^3 \bar{G}_{5X} \mathcal{\bar{K}}_3 \;, \nonumber \\
 \bar{F}_{g^{\tau\tau}} = \ & \mathcal{G}_5\mathcal{\bar{K}}_3 + \frac{1}{2}\dot{\bar{\Phi}}^2 \mathcal{J}_5 \mathcal{\bar{K}}_2 - \sqrt{-\bar{g}^{\tau\tau}}\dot{\bar{\Phi}}^3 \bar{G}_{5X} \mathcal{\bar{R}}_3 + \bar{F}_{4X} \dot{\bar{\Phi}}^2 {}^{(3)}\!\bar{R} \;, \nonumber \\
 \bar{F}_{K} = \ &\frac{1}{6} \sqrt{-\bar{g}^{\tau\tau}}\dot{\bar{\Phi}} \bar{F}_5 {}^{(3)}\!\bar{R} - \frac{1}{3} (-\bar{g}^{\tau\tau})^{3/2} \dot{\bar{\Phi}}^3 \bar{G}_{5X} \mathcal{\bar{K}}_2 + \frac{2}{3} \bar{g}^{\tau\tau} \dot{\bar{\Phi}}^2 \bar{K} \bar{G}_{5\Phi}  \;, \nonumber \\ 
  \bar{F}_{\sigma^\mu_\nu} = \ & - \sqrt{-\bar{g}^{\tau\tau}} \dot{\bar{\Phi}} \bar{F}_5 \bar{r}^\nu_\mu + \frac{2}{3} (-\bar{g}^{\tau\tau})^{3/2} \dot{\bar{\Phi}}^3 \bar{G}_{5X} \mathcal{S}^\nu_\mu - \bar{g}^{\tau\tau} \dot{\bar{\Phi}}^2 \bar{G}_{5\Phi}\bar{\sigma}^\nu_\mu \nonumber \\ 
  \bar{F}_{{}^{(3)}\!R} = \ & \frac{1}{6} \sqrt{-\bar{g}^{\tau\tau}} \dot{\bar{\Phi}} \bar{K} \bar{F}_5 + \bar{F}_4 \;, \quad \bar{F}_{r^\mu_\nu} = - \sqrt{-\bar{g}^{\tau\tau}} \dot{\bar{\Phi}} \bar{F}_5 \bar{\sigma}_\mu^\nu \;, \nonumber \\
 \bar{F}_{g^{\tau\tau} g^{\tau\tau}} = \ & \bigg(-\frac{\dot{\bar{\Phi}}^3 \bar{G}_{5X}}{4\sqrt{-\bar{g}^{\tau\tau}}}  + \sqrt{-\bar{g}^{\tau\tau}} \dot{\bar{\Phi}}^5 \bar{G}_{5XX} - \frac{1}{3}(-\bar{g}^{\tau\tau})^{3/2} \dot{\bar{\Phi}}^7 \bar{G}_{5XXX} \bigg) \mathcal{\bar{K}}_3 + \bar{F}_{4XX}\dot{\bar{\Phi}}^4 {}^{(3)}\!\bar{R}  \\ 
 &+ \bigg(\frac{\dot{\bar{\Phi}}^3 \bar{G}_{5X}}{2 \sqrt{-\bar{g}^{\tau\tau}}}  - \sqrt{-\bar{g}^{\tau\tau}} \dot{\bar{\Phi}}^5 \bar{G}_{5 XX} \bigg)\mathcal{\bar{R}}_3 + \frac{1}{2}\dot{\bar{\Phi}}^4  (2 \bar{G}_{5\Phi X} + \bar{g}^{\tau\tau} \dot{\bar{\Phi}}^2 \bar{G}_{5\Phi XX}) \mathcal{\bar{K}}_2 \;, \nonumber \\
 \bar{F}_{g^{\tau\tau} K} = \ & \frac{1}{6}\sqrt{-\bar{g}^{\tau\tau}} \dot{\bar{\Phi}}^3 \bar{G}_{5X}  {}^{(3)}\!\bar{R} + \mathcal{G}_5 \mathcal{\bar{K}}_2 + \frac{2}{3}\dot{\bar{\Phi}}^2 \bar{K} \mathcal{J}_5 \;, \nonumber \\
 \bar{F}_{g^{\tau\tau}\sigma^\mu_\nu} = \ &-\sqrt{-\bar{g}^{\tau\tau}}\dot{\bar{\Phi}}^3 \bar{G}_{5X} \bar{r}^\nu_\mu - 2 \mathcal{G}_5 \mathcal{S}_\mu^\nu - \dot{\bar{\Phi}}^2 \bar{\sigma}^\nu_\mu \mathcal{J}_5 \;,  \nonumber \\
 \bar{F}_{g^{\tau\tau}r^\mu_\nu} = \ &-\sqrt{-\bar{g}^{\tau\tau}}\dot{\bar{\Phi}}^3 \bar{G}_{5X} \bar{\sigma}^\nu_\mu \;, \quad \bar{F}_{g^{\tau\tau} {}^{(3)}\!R} =  \frac{1}{6}\sqrt{-\bar{g}^{\tau\tau}}\dot{\bar{\Phi}}^3 \bar{K} \bar{G}_{5X} + \bar{F}_{4X}\dot{\bar{\Phi}}^2 \;, \nonumber \\ 
\bar{F}_{KK} = \ & -\frac{4}{9} (-\bar{g}^{\tau\tau})^{3/2} \dot{\bar{\Phi}}^3 \bar{K} \bar{G}_{5X} +  \frac{2}{3} \bar{g}^{\tau\tau} \dot{\bar{\Phi}}^2 \bar{G}_{5\Phi} \;, \quad
 \bar{F}_{K {}^{(3)}\!R} =  \frac{1}{6} \sqrt{-\bar{g}^{\tau\tau}} \dot{\bar{\Phi}} \bar{F}_5 \;, \nonumber \\
  \bar{F}_{K \sigma^\mu_\nu} = \ & \frac{2}{3}  (-\bar{g}^{\tau\tau})^{3/2} \dot{\bar{\Phi}}^3 \bar{G}_{5X}  \bar{\sigma}^\nu_\mu \;, \quad 
  \bar{F}_{\sigma^2} = \frac{2}{3} (-\bar{g}^{\tau\tau})^{3/2} \dot{\bar{\Phi}}^3 \bar{K} \bar{G}_{5X} - \bar{g}^{\tau\tau} \dot{\bar{\Phi}}^2 \bar{G}_{5\Phi} \;, \quad
    \bar{F}_{\sigma r} = -\sqrt{-\bar{g}^{\tau\tau}}\dot{\bar{\Phi}} \bar{F}_5 \;, \nonumber
 \end{align} 
where $\mathcal{S}_\mu^\nu \equiv \bar{K} \bar{\sigma}^\nu_\mu - 3 \bar{\sigma}^\alpha_\mu \bar{\sigma}^\nu_\alpha$, $\mathcal{G}_5 \equiv (3 \sqrt{-\bar{g}^{\tau\tau}} \dot{\bar{\Phi}}^3 \bar{G}_{5X} - 2(-\bar{g}^{\tau\tau})^{3/2}\dot{\bar{\Phi}}^5 \bar{G}_{5XX})/6$ and $\mathcal{J}_5 \equiv \bar{G}_{5\Phi} + \bar{g}^{\tau\tau} \dot{\bar{\Phi}}^2 \bar{G}_{5\Phi X}$.

The EFT parameters are therefore given by
\begin{align}
M_\star^2 f = \ & \frac{1}{3}\sqrt{-\bar{g}^{\tau\tau}} \dot{\bar{\Phi}} \bar{K} \bar{F}_5 + 2 \bar{F}_4 \;, \quad c = -\bar{F}_{g^{\tau\tau}}\;, \nonumber \\ 
\Lambda = \ &\bigg( \bar{F}_{4} + \frac{1}{6} \sqrt{-\bar{g}^{\tau\tau}}\dot{\bar{\Phi}} \bar{K} \bar{F}_5 - \frac{1}{3} (-\bar{g}^{\tau\tau})^{3/2}  \dot{\bar{\Phi}}^3 \bar{K} \bar{G}_{5 X} + \frac{1}{2} (-\bar{g}^{\tau\tau})^2 \dot{\bar{\Phi}}^4 \bar{G}_{5 X \Phi} + \bar{g}^{\tau\tau} \dot{\bar{\Phi}}^2 \bar{G}_{5\Phi}\bigg) \mathcal{\bar{K}}_2  \;, \nonumber \\
& + \bigg(-\sqrt{-\bar{g}^{\tau\tau}}\dot{\bar{\Phi}}\bar{F}_5 + (-\bar{g}^{\tau\tau})^{3/2} \dot{\bar{\Phi}}^3 \bar{G}_{5X} \bigg) \mathcal{\bar{R}}_3 +  \frac{2}{3} (-\bar{g}^{\tau\tau})^{3/2} \dot{\bar{\Phi}}^3 \bar{G}_{5X} \bar{\sigma}^\mu_\nu \mathcal{S}^\nu_\mu  \nonumber \\ 
&- \frac{1}{6} \dot{\bar{\Phi}}^3 ((-\bar{g}^{\tau\tau})^{3/2} \bar{G}_{5X} - 2 (-\bar{g}^{\tau\tau})^{5/2} \dot{\bar{\Phi}}^2 \bar{G}_{5XX} )\mathcal{\bar{K}}_3 + \bar{g}^{\tau\tau}\dot{\bar{\Phi}}^2 \bar{F}_{4X} {}^{(3)}\!\bar{R} \nonumber \\ 
 \beta = \ & - \frac{1}{3} \sqrt{-\bar{g}^{\tau\tau}} \dot{\bar{\Phi}}\bar{F}_5 \bigg(\frac{1}{2} {}^{(3)}\!\bar{R} + \frac{2}{3} \bar{K}^2 \bigg) + \frac{1}{3} (-\bar{g}^{\tau\tau})^{3/2} \dot{\bar{\Phi}}^3 \bar{G}_{5X} \mathcal{\bar{K}}_2 - \frac{4}{3} \bar{g}^{\tau\tau} \dot{\bar{\Phi}}^2 \bar{K} \bar{G}_{5\Phi}  + \frac{2}{3} \bar{g}^{\tau\tau} \dot{\bar{\Phi}}^2 \bar{K} \bar{F}_{5\Phi}\;, \nonumber \\
\alpha^\mu_\nu = \ & \sqrt{-\bar{g}^{\tau\tau}} \dot{\bar{\Phi}} \bar{F}_5 \bigg(\bar{r}^\mu_\nu + \frac{1}{3} \bar{K} \bar{\sigma}^\mu_\nu \bigg) - \bar{g}^{\tau\tau} \dot{\bar{\Phi}}^2 (\bar{F}_{5 \Phi} - 2 \bar{G}_{5\Phi})\bar{\sigma}^\mu_\nu - \frac{2}{3} (-\bar{g}^{\tau\tau})^{3/2} \dot{\bar{\Phi}}^3 \bar{G}_{5X} \mathcal{S}^\mu_\nu \;, \nonumber \\
 \gamma^\mu_\nu = \ & \sqrt{-\bar{g}^{\tau\tau}} \dot{\bar{\Phi}} \bar{F}_5 \bar{\sigma}^\mu_\nu \;, \quad 
m_2^4 = \bar{F}_{g^{\tau\tau} g^{\tau\tau}} \;, \\ 
 M_1^3 = \ & \frac{1}{3}\sqrt{-\bar{g}^{\tau\tau}} \dot{\bar{\Phi}}^3 \bar{G}_{5X} {}^{(3)}\!\bar{R}  + 2 \mathcal{G}_5( \mathcal{\bar{K}}_2 - 2 \bar{\sigma}^\mu_\nu \bar{\sigma}^\nu_\mu )  + \frac{4}{3} \dot{\bar{\Phi}}^2 \bar{K} \mathcal{J}_5 \nonumber  \\ 
M_2^2 = \ & -\frac{2}{3} (-\bar{g}^{\tau\tau})^{3/2} \dot{\bar{\Phi}}^3 \bar{K} \bar{G}_{5X} + 2 \bar{g}^{\tau\tau} \dot{\bar{\Phi}}^2 \bar{G}_{5\Phi} + \frac{1}{3}\sqrt{-\bar{g}^{\tau\tau}} \dot{\bar{\Phi}} \bar{K} \bar{F}_5 - \bar{g}^{\tau\tau} \dot{\bar{\Phi}}^2 \bar{F}_{5\Phi} \;, \nonumber \\ 
M_3^2 = \ & \frac{2}{3} (-\bar{g}^{\tau\tau})^{3/2} \dot{\bar{\Phi}}^3 \bar{K} \bar{G}_{5X}  - 2 \bar{g}^{\tau\tau} \dot{\bar{\Phi}}^2 \bar{G}_{5\Phi} - \frac{1}{3} \sqrt{-\bar{g}^{\tau\tau}} \dot{\bar{\Phi}} \bar{K} \bar{F}_5 + \bar{g}^{\tau\tau} \dot{\bar{\Phi}}^2 \bar{F}_{5\Phi}  \nonumber \;, \\
M_4 = \ & \sqrt{-\bar{g}^{\tau\tau}} \dot{\bar{\Phi}} \bar{F}_5 \;, \quad M_5 = - 2\sqrt{-\bar{g}^{\tau\tau}} \dot{\bar{\Phi}} \bar{F}_5 \;, \quad \mu_1^2 = \frac{1}{3}\sqrt{-\bar{g}^{\tau\tau}} \dot{\bar{\Phi}}^3 \bar{K} \bar{G}_{5X} +  \dot{\bar{\Phi}}^2 \bar{G}_{5\Phi} - \frac{1}{2} \dot{\bar{\Phi}}^2 \bar{F}_{5\Phi} \nonumber \;, \\
\lambda^\mu_{1 \ \nu} = \ & -2 \sqrt{-\bar{g}^{\tau\tau}} \dot{\bar{\Phi}}^3 \bar{G}_{5X} \bar{r}^\mu_\nu - 4 \mathcal{G}_5 \mathcal{S}^\mu_\nu - 2 \dot{\bar{\Phi}}^2 \bar{\sigma}^\mu_\nu \mathcal{J}_5 \;, \nonumber \\
\lambda^\mu_{2 \ \nu} = \ & -2 \sqrt{-\bar{g}^{\tau\tau}} \dot{\bar{\Phi}}^3 \bar{G}_{5X}\bar{\sigma}^\mu_\nu \;, \quad \lambda^\mu_{3 \ \nu} = \frac{4}{3} (-\bar{g}^{\tau\tau})^{3/2} \dot{\bar{\Phi}}^3 \bar{G}_{5X}\bar{\sigma}^\mu_\nu \;. \nonumber
\end{align}
Note that the auxiliary function $F_5$ in the expressions above can be written in terms of $G_{5X}$ as
\begin{align}
F_5 = \frac{1}{\sqrt{-X}} \int dX' \sqrt{-X'} G_{5X'} \;,
\end{align}
and $F_4 = X(G_{5\Phi} - F_{5\Phi})/2$.
{}
\bibliographystyle{utphys}
\bibliography{bib_v4}

\providecommand{\href}[2]{#2}\begingroup\raggedright\begin{thebibliography}{10}

\bibitem{jordan1955schwerkraft}
P.~Jordan, {\em Schwerkraft und Weltall: Grundlagen d. theoret. Kosmologie. Mit
  13 Abb}.
\newblock Die Wissenschaft. Vieweg, 1955.

\bibitem{Brans:1961sx}
C.~Brans and R.~Dicke, ``{Mach's principle and a relativistic theory of
  gravitation},'' {\em Phys.Rev.} {\bf 124} (1961) 925--935.

\bibitem{Bergmann:1968ve}
P.~G. Bergmann, ``{Comments on the scalar tensor theory},'' {\em Int. J. Theor.
  Phys.} {\bf 1} (1968) 25--36.

\bibitem{Wagoner:1970vr}
R.~V. Wagoner, ``{Scalar tensor theory and gravitational waves},'' {\em Phys.
  Rev. D} {\bf 1} (1970) 3209--3216.

\bibitem{Horndeski:1974wa}
G.~W. Horndeski, ``{Second-order scalar-tensor field equations in a
  four-dimensional space},'' {\em Int.J.Theor.Phys.} {\bf 10} (1974) 363--384.

\bibitem{Langlois:2015cwa}
D.~Langlois and K.~Noui, ``{Degenerate higher derivative theories beyond
  Horndeski: evading the Ostrogradski instability},'' {\em JCAP} {\bf 1602}
  (2016), no.~02 034, \href{https://arxiv.org/abs/1510.06930}{{\tt
  1510.06930}}.

\bibitem{Langlois:2018dxi}
D.~Langlois, ``{Dark energy and modified gravity in degenerate higher-order
  scalar\textendash{}tensor (DHOST) theories: A review},'' {\em Int. J. Mod.
  Phys. D} {\bf 28} (2019), no.~05 1942006,
  \href{https://arxiv.org/abs/1811.06271}{{\tt 1811.06271}}.

\bibitem{ArkaniHamed:2003uy}
N.~Arkani-Hamed, H.-C. Cheng, M.~A. Luty, and S.~Mukohyama, ``{Ghost
  condensation and a consistent infrared modification of gravity},'' {\em JHEP}
  {\bf 05} (2004) 074, \href{https://arxiv.org/abs/hep-th/0312099}{{\tt
  hep-th/0312099}}.

\bibitem{Arkani-Hamed:2003juy}
N.~Arkani-Hamed, P.~Creminelli, S.~Mukohyama, and M.~Zaldarriaga, ``{Ghost
  inflation},'' {\em JCAP} {\bf 04} (2004) 001,
  \href{https://arxiv.org/abs/hep-th/0312100}{{\tt hep-th/0312100}}.

\bibitem{Creminelli:2006xe}
P.~Creminelli, M.~A. Luty, A.~Nicolis, and L.~Senatore, ``{Starting the
  Universe: Stable Violation of the Null Energy Condition and Non-standard
  Cosmologies},'' {\em JHEP} {\bf 0612} (2006) 080,
  \href{https://arxiv.org/abs/hep-th/0606090}{{\tt hep-th/0606090}}.

\bibitem{Cheung:2007st}
C.~Cheung, P.~Creminelli, A.~L. Fitzpatrick, J.~Kaplan, and L.~Senatore, ``{The
  Effective Field Theory of Inflation},'' {\em JHEP} {\bf 0803} (2008) 014,
  \href{https://arxiv.org/abs/0709.0293}{{\tt 0709.0293}}.

\bibitem{Creminelli:2005hu}
P.~Creminelli, A.~Nicolis, L.~Senatore, M.~Tegmark, and M.~Zaldarriaga,
  ``{Limits on non-gaussianities from wmap data},'' {\em JCAP} {\bf 05} (2006)
  004, \href{https://arxiv.org/abs/astro-ph/0509029}{{\tt astro-ph/0509029}}.

\bibitem{Senatore:2009gt}
L.~Senatore, K.~M. Smith, and M.~Zaldarriaga, ``{Non-Gaussianities in Single
  Field Inflation and their Optimal Limits from the WMAP 5-year Data},'' {\em
  JCAP} {\bf 01} (2010) 028, \href{https://arxiv.org/abs/0905.3746}{{\tt
  0905.3746}}.

\bibitem{Gubitosi:2012hu}
G.~Gubitosi, F.~Piazza, and F.~Vernizzi, ``{The Effective Field Theory of Dark
  Energy},'' {\em JCAP} {\bf 1302} (2013) 032,
  \href{https://arxiv.org/abs/1210.0201}{{\tt 1210.0201}}.

\bibitem{Creminelli:2017sry}
P.~Creminelli and F.~Vernizzi, ``{Dark Energy after GW170817 and GRB170817A},''
  {\em Phys. Rev. Lett.} {\bf 119} (2017), no.~25 251302,
  \href{https://arxiv.org/abs/1710.05877}{{\tt 1710.05877}}.

\bibitem{Ezquiaga:2017ekz}
J.~M. Ezquiaga and M.~Zumalacrregui, ``{Dark Energy After GW170817: Dead Ends
  and the Road Ahead},'' {\em Phys. Rev. Lett.} {\bf 119} (2017), no.~25
  251304, \href{https://arxiv.org/abs/1710.05901}{{\tt 1710.05901}}.

\bibitem{Baker:2017hug}
T.~Baker, E.~Bellini, P.~G. Ferreira, M.~Lagos, J.~Noller, and I.~Sawicki,
  ``{Strong constraints on cosmological gravity from GW170817 and GRB
  170817A},'' {\em Phys. Rev. Lett.} {\bf 119} (2017), no.~25 251301,
  \href{https://arxiv.org/abs/1710.06394}{{\tt 1710.06394}}.

\bibitem{Creminelli:2018xsv}
P.~Creminelli, M.~Lewandowski, G.~Tambalo, and F.~Vernizzi, ``{Gravitational
  Wave Decay into Dark Energy},'' {\em JCAP} {\bf 1812} (2018), no.~12 025,
  \href{https://arxiv.org/abs/1809.03484}{{\tt 1809.03484}}.

\bibitem{Creminelli:2019nok}
P.~Creminelli, G.~Tambalo, F.~Vernizzi, and V.~Yingcharoenrat, ``{Resonant
  Decay of Gravitational Waves into Dark Energy},'' {\em JCAP} {\bf 10} (2019)
  072, \href{https://arxiv.org/abs/1906.07015}{{\tt 1906.07015}}.

\bibitem{Creminelli:2019kjy}
P.~Creminelli, G.~Tambalo, F.~Vernizzi, and V.~Yingcharoenrat, ``{Dark-Energy
  Instabilities induced by Gravitational Waves},'' {\em JCAP} {\bf 05} (2020)
  002, \href{https://arxiv.org/abs/1910.14035}{{\tt 1910.14035}}.

\bibitem{Cheng:2006us}
H.-C. Cheng, M.~A. Luty, S.~Mukohyama, and J.~Thaler, ``{Spontaneous Lorentz
  breaking at high energies},'' {\em JHEP} {\bf 05} (2006) 076,
  \href{https://arxiv.org/abs/hep-th/0603010}{{\tt hep-th/0603010}}.

\bibitem{Aoki:2021wew}
K.~Aoki, M.~A. Gorji, S.~Mukohyama, and K.~Takahashi, ``{The effective field
  theory of vector-tensor theories},'' {\em JCAP} {\bf 01} (2022), no.~01 059,
  \href{https://arxiv.org/abs/2111.08119}{{\tt 2111.08119}}.

\bibitem{Mukohyama:2005rw}
S.~Mukohyama, ``{Black holes in the ghost condensate},'' {\em Phys. Rev. D}
  {\bf 71} (2005) 104019, \href{https://arxiv.org/abs/hep-th/0502189}{{\tt
  hep-th/0502189}}.

\bibitem{Babichev:2013cya}
E.~Babichev and C.~Charmousis, ``{Dressing a black hole with a time-dependent
  Galileon},'' {\em JHEP} {\bf 08} (2014) 106,
  \href{https://arxiv.org/abs/1312.3204}{{\tt 1312.3204}}.

\bibitem{Kobayashi:2014eva}
T.~Kobayashi and N.~Tanahashi, ``{Exact black hole solutions in shift symmetric
  scalar\textendash{}tensor theories},'' {\em PTEP} {\bf 2014} (2014) 073E02,
  \href{https://arxiv.org/abs/1403.4364}{{\tt 1403.4364}}.

\bibitem{BenAchour:2018dap}
J.~Ben~Achour and H.~Liu, ``{Hairy Schwarzschild-(A)dS black hole solutions in
  degenerate higher order scalar-tensor theories beyond shift symmetry},'' {\em
  Phys. Rev. D} {\bf 99} (2019), no.~6 064042,
  \href{https://arxiv.org/abs/1811.05369}{{\tt 1811.05369}}.

\bibitem{Motohashi:2019sen}
H.~Motohashi and M.~Minamitsuji, ``{Exact black hole solutions in
  shift-symmetric quadratic degenerate higher-order scalar-tensor theories},''
  {\em Phys. Rev. D} {\bf 99} (2019), no.~6 064040,
  \href{https://arxiv.org/abs/1901.04658}{{\tt 1901.04658}}.

\bibitem{Charmousis:2019vnf}
C.~Charmousis, M.~Crisostomi, R.~Gregory, and N.~Stergioulas, ``{Rotating Black
  Holes in Higher Order Gravity},'' {\em Phys. Rev. D} {\bf 100} (2019), no.~8
  084020, \href{https://arxiv.org/abs/1903.05519}{{\tt 1903.05519}}.

\bibitem{BenAchour:2020wiw}
J.~Ben~Achour, H.~Liu, and S.~Mukohyama, ``{Hairy black holes in DHOST
  theories: Exploring disformal transformation as a solution-generating
  method},'' {\em JCAP} {\bf 02} (2020) 023,
  \href{https://arxiv.org/abs/1910.11017}{{\tt 1910.11017}}.

\bibitem{BenAchour:2020fgy}
J.~Ben~Achour, H.~Liu, H.~Motohashi, S.~Mukohyama, and K.~Noui, ``{On rotating
  black holes in DHOST theories},'' {\em JCAP} {\bf 11} (2020) 001,
  \href{https://arxiv.org/abs/2006.07245}{{\tt 2006.07245}}.

\bibitem{Anson:2020trg}
T.~Anson, E.~Babichev, C.~Charmousis, and M.~Hassaine, ``{Disforming the Kerr
  metric},'' {\em JHEP} {\bf 01} (2021) 018,
  \href{https://arxiv.org/abs/2006.06461}{{\tt 2006.06461}}.

\bibitem{deRham:2019gha}
C.~de~Rham and J.~Zhang, ``{Perturbations of stealth black holes in degenerate
  higher-order scalar-tensor theories},'' {\em Phys. Rev. D} {\bf 100} (2019),
  no.~12 124023, \href{https://arxiv.org/abs/1907.00699}{{\tt 1907.00699}}.

\bibitem{Motohashi:2019ymr}
H.~Motohashi and S.~Mukohyama, ``{Weakly-coupled stealth solution in scordatura
  degenerate theory},'' {\em JCAP} {\bf 01} (2020) 030,
  \href{https://arxiv.org/abs/1912.00378}{{\tt 1912.00378}}.

\bibitem{Gorji:2020bfl}
M.~A. Gorji, H.~Motohashi, and S.~Mukohyama, ``{Stealth dark energy in
  scordatura DHOST theory},'' {\em JCAP} {\bf 03} (2021) 081,
  \href{https://arxiv.org/abs/2009.11606}{{\tt 2009.11606}}.

\bibitem{Gorji:2021isn}
M.~A. Gorji, H.~Motohashi, and S.~Mukohyama, ``{Inflation with
  0~\ensuremath{\leq}~c $_{s}$~\ensuremath{\leq}~1},'' {\em JCAP} {\bf 02}
  (2022), no.~02 030, \href{https://arxiv.org/abs/2110.10731}{{\tt
  2110.10731}}.

\bibitem{Hui:2012qt}
L.~Hui and A.~Nicolis, ``{No-Hair Theorem for the Galileon},'' {\em Phys. Rev.
  Lett.} {\bf 110} (2013) 241104, \href{https://arxiv.org/abs/1202.1296}{{\tt
  1202.1296}}.

\bibitem{Saravani:2019xwx}
M.~Saravani and T.~P. Sotiriou, ``{Classification of shift-symmetric Horndeski
  theories and hairy black holes},'' {\em Phys. Rev. D} {\bf 99} (2019), no.~12
  124004, \href{https://arxiv.org/abs/1903.02055}{{\tt 1903.02055}}.

\bibitem{Babichev:2016rlq}
E.~Babichev, C.~Charmousis, and A.~Leh\'ebel, ``{Black holes and stars in
  Horndeski theory},'' {\em Class. Quant. Grav.} {\bf 33} (2016), no.~15
  154002, \href{https://arxiv.org/abs/1604.06402}{{\tt 1604.06402}}.

\bibitem{Antoniou:2017acq}
G.~Antoniou, A.~Bakopoulos, and P.~Kanti, ``{Evasion of No-Hair Theorems and
  Novel Black-Hole Solutions in Gauss-Bonnet Theories},'' {\em Phys. Rev.
  Lett.} {\bf 120} (2018), no.~13 131102,
  \href{https://arxiv.org/abs/1711.03390}{{\tt 1711.03390}}.

\bibitem{Franciolini:2018uyq}
G.~Franciolini, L.~Hui, R.~Penco, L.~Santoni, and E.~Trincherini, ``{Effective
  Field Theory of Black Hole Quasinormal Modes in Scalar-Tensor Theories},''
  {\em JHEP} {\bf 02} (2019) 127, \href{https://arxiv.org/abs/1810.07706}{{\tt
  1810.07706}}.

\bibitem{Cusin:2017mzw}
G.~Cusin, M.~Lewandowski, and F.~Vernizzi, ``{Nonlinear Effective Theory of
  Dark Energy},'' {\em JCAP} {\bf 1804} (2018), no.~04 061,
  \href{https://arxiv.org/abs/1712.02782}{{\tt 1712.02782}}.

\bibitem{Arkani-Hamed:2003pdi}
N.~Arkani-Hamed, H.-C. Cheng, M.~A. Luty, and S.~Mukohyama, ``{Ghost
  condensation and a consistent infrared modification of gravity},'' {\em JHEP}
  {\bf 05} (2004) 074, \href{https://arxiv.org/abs/hep-th/0312099}{{\tt
  hep-th/0312099}}.

\bibitem{Deffayet:2009mn}
C.~Deffayet, S.~Deser, and G.~Esposito-Farese, ``{Generalized Galileons: All
  scalar models whose curved background extensions maintain second-order field
  equations and stress-tensors},'' {\em Phys. Rev.} {\bf D80} (2009) 064015,
  \href{https://arxiv.org/abs/0906.1967}{{\tt 0906.1967}}.

\bibitem{Gleyzes:2013ooa}
J.~Gleyzes, D.~Langlois, F.~Piazza, and F.~Vernizzi, ``{Essential Building
  Blocks of Dark Energy},'' {\em JCAP} {\bf 1308} (2013) 025,
  \href{https://arxiv.org/abs/1304.4840}{{\tt 1304.4840}}.

\bibitem{Mukohyama:2022skv}
S.~Mukohyama, K.~Takahashi, and V.~Yingcharoenrat, ``{Generalized Regge-Wheeler
  Equation from Effective Field Theory of Black Hole Perturbations with a
  Timelike Scalar Profile},'' {\em \textit{in preparation}}.

\bibitem{Lemaitre:1933gd}
G.~Lemaitre, ``{The expanding universe},'' {\em Annales Soc. Sci. Bruxelles A}
  {\bf 53} (1933) 51--85.

\bibitem{Takahashi:2021bml}
K.~Takahashi and H.~Motohashi, ``{Black hole perturbations in DHOST theories:
  master variables, gradient instability, and strong coupling},'' {\em JCAP}
  {\bf 08} (2021) 013, \href{https://arxiv.org/abs/2106.07128}{{\tt
  2106.07128}}.

\bibitem{Tomikawa:2021pca}
K.~Tomikawa and T.~Kobayashi, ``{Perturbations and quasinormal modes of black
  holes with time-dependent scalar hair in shift-symmetric scalar-tensor
  theories},'' {\em Phys. Rev. D} {\bf 103} (2021), no.~8 084041,
  \href{https://arxiv.org/abs/2101.03790}{{\tt 2101.03790}}.

\bibitem{Hui:2020xxx}
L.~Hui, A.~Joyce, R.~Penco, L.~Santoni, and A.~R. Solomon, ``{Static response
  and Love numbers of Schwarzschild black holes},'' {\em JCAP} {\bf 04} (2021)
  052, \href{https://arxiv.org/abs/2010.00593}{{\tt 2010.00593}}.

\bibitem{Charalambous:2021mea}
P.~Charalambous, S.~Dubovsky, and M.~M. Ivanov, ``{On the Vanishing of Love
  Numbers for Kerr Black Holes},'' {\em JHEP} {\bf 05} (2021) 038,
  \href{https://arxiv.org/abs/2102.08917}{{\tt 2102.08917}}.

\bibitem{Cardoso:2018ptl}
V.~Cardoso, M.~Kimura, A.~Maselli, and L.~Senatore, ``{Black Holes in an
  Effective Field Theory Extension of General Relativity},'' {\em Phys. Rev.
  Lett.} {\bf 121} (2018), no.~25 251105,
  \href{https://arxiv.org/abs/1808.08962}{{\tt 1808.08962}}.

\bibitem{Bernard:2019yfz}
L.~Bernard, ``{Dipolar tidal effects in scalar-tensor theories},'' {\em Phys.
  Rev. D} {\bf 101} (2020), no.~2 021501,
  \href{https://arxiv.org/abs/1906.10735}{{\tt 1906.10735}}.

\bibitem{Hui:2021cpm}
L.~Hui, A.~Podo, L.~Santoni, and E.~Trincherini, ``{Effective Field Theory for
  the Perturbations of a Slowly Rotating Black Hole},''
  \href{https://arxiv.org/abs/2111.02072}{{\tt 2111.02072}}.

\bibitem{Takahashi:2019oxz}
K.~Takahashi, H.~Motohashi, and M.~Minamitsuji, ``{Linear stability analysis of
  hairy black holes in quadratic degenerate higher-order scalar-tensor
  theories: Odd-parity perturbations},'' {\em Phys. Rev. D} {\bf 100} (2019),
  no.~2 024041, \href{https://arxiv.org/abs/1904.03554}{{\tt 1904.03554}}.

\bibitem{Takahashi:2020hso}
K.~Takahashi and H.~Motohashi, ``{General Relativity solutions with stealth
  scalar hair in quadratic higher-order scalar-tensor theories},'' {\em JCAP}
  {\bf 06} (2020) 034, \href{https://arxiv.org/abs/2004.03883}{{\tt
  2004.03883}}.

\end{thebibliography}\endgroup

\end{document}